\documentclass[preprintnumbers,article,amsmath,amssymb,floatfix,10pt,prd,onecolumn,
superscriptaddress,nofootinbib]{revtex4-2}
\usepackage{bm}
\usepackage{amsfonts}
\usepackage{latexsym}
\usepackage[latin1]{inputenc}
\usepackage{graphicx}
\usepackage{amsmath}
\usepackage{palatino}
\usepackage{mathpazo}
\usepackage{textcomp}
\usepackage{float}
\usepackage{booktabs}
\usepackage{dcolumn}
\usepackage{ragged2e}
\usepackage{hyperref}
\hypersetup{colorlinks,citecolor=blue}
\hypersetup{colorlinks=true,linkcolor=red,filecolor=magenta,    urlcolor=blue}
\usepackage{amsmath}
\usepackage{xcolor}
\usepackage{orcidlink}
\usepackage{epsfig}
\usepackage{caption}
\usepackage{subcaption}
\usepackage{commath}
\def\jnl@style{\it}
\def\aaref@jnl#1{{\jnl@style#1}}

\def\aaref@jnl#1{{\jnl@style#1}}

\def\aj{\aaref@jnl{AJ}}                   
\def\apj{\aaref@jnl{ApJ}}                 
\def\apjl{\aaref@jnl{ApJ}}                
\def\apjs{\aaref@jnl{ApJS}}               
\def\apss{\aaref@jnl{Ap\&SS}}             
\def\aap{\aaref@jnl{A\&A}}                
\def\aapr{\aaref@jnl{A\&A~Rev.}}          
\def\aaps{\aaref@jnl{A\&AS}}              
\def\mnras{\aaref@jnl{Mon.~Not.~Roy.~Astron.~Soc.}}             
\def\prd{\aaref@jnl{Phys.~Rev.~D}}        
\def\prc{\aaref@jnl{Phys.~Rev.~C}}  
\def\prl{\aaref@jnl{Phys.~Rev.~Lett.}}    
\def\qjras{\aaref@jnl{QJRAS}}             
\def\skytel{\aaref@jnl{S\&T}}             
\def\ssr{\aaref@jnl{Space~Sci.~Rev.}}     
\def\zap{\aaref@jnl{ZAp}}                 
\def\nat{\aaref@jnl{Nature}}              
\def\aplett{\aaref@jnl{Astrophys.~Lett.}} 
\def\apspr{\aaref@jnl{Astrophys.~Space~Phys.~Res.}} 
\def\physrep{\aaref@jnl{Phys.~Rep.}}      
\def\physscr{\aaref@jnl{Phys.~Scr}}       
\def\commat{\aaref@jnl{Comm.~Math.~Phys.}}              
\def\science{\aaref@jnl{Science}}               
\def\cqg{\aaref@jnl{Classical Quant.~Grav.}}            
\def\jpcs{\aaref@jnl{JPCS}}                                     
\def\ijmpd{\aaref@jnl{Int.~J.~Mod.~Phys.~D}}                    
\def\grg{\aaref@jnl{Gen.~Relat.~Gravit.}}               
\def\rpp{\aaref@jnl{Rep.~Prog.~Phys.}}          
\def\npa{\aaref@jnl{Nucl.~Phys.~A}}        
\def\lrr{\aaref@jnl{Living Rev.~Rel.}}                   
\def\jcap{\aaref@jnl{J.~Cosmology Astropart.~Phys.}}    
\def\rmp{\aaref@jnl{Rev.~Mod.~Phys.}}   
\def\epjc{\aaref@jnl{Eur.~Phys.~J.~C}} 
\def\plb{\aaref@jnl{~Phy.~Lett.~B}} 
\def\mpla{\aaref@jnl{Mod.~Phy.~Lett.~A}} 
\def\arxiv{\aaref@jnl{arxiv.org}}

\allowdisplaybreaks[1]

\begin{document}
\color{black}       
%
\title{Correction to Lagrangian for Bouncing Cosmologies in $f(Q)$ Gravity}

\author{Gaurav N. Gadbail\orcidlink{0000-0003-0684-9702}}
\email{gauravgadbail6@gmail.com}
\affiliation{Department of Mathematics, Birla Institute of Technology and
Science-Pilani,\\ Hyderabad Campus, Hyderabad-500078, India.}

\author{Ameya Kolhatkar\orcidlink{0000-0002-9553-1220}}
\email{kolhatkarameya1996@gmail.com}
\affiliation{Department of Mathematics, Birla Institute of Technology and
Science-Pilani,\\ Hyderabad Campus, Hyderabad-500078, India.}

\author{Sanjay Mandal\orcidlink{0000-0003-2570-2335}}
\email{sanjaymandal960@gmail.com}
\affiliation{Faculty of Mathematics \& Computer Science, Transilvania University of Brasov, Eroilor 29, Brasov, Romania}

\author{P.K. Sahoo\orcidlink{0000-0003-2130-8832}}
\email{pksahoo@hyderabad.bits-pilani.ac.in}
\affiliation{Department of Mathematics, Birla Institute of Technology and
Science-Pilani,\\ Hyderabad Campus, Hyderabad-500078, India.}
\affiliation{Faculty of Mathematics \& Computer Science, Transilvania University of Brasov, Eroilor 29, Brasov, Romania}
%

\begin{abstract}
Symmetric teleparallel gravity offers to reformulate the gravitational formalism without the presence of curvature and torsion with the help of non-metricity tensors. Interestingly, Symmetric teleparallel gravity can be formulated equivalently to teleparallel gravity or general relativity for an appropriate setup. In this study, our aim lies in exploring the bouncing cosmologies as an alternative to the initial singularity of the Universe in the background of modified symmetric teleparallel gravity. To explore this, we adopt the reconstruction technique to present the possible reconstructed Lagrangian for various cosmological bouncing solutions in a flat Friedmann-Lema\^itre-Robertson-Walker spacetime with a perfect fluid matter distribution. We study the reconstructed gravitational Lagrangians, which are capable of reproducing analytical solutions for \textit{symmetric bounce}, \textit{super-bounce}, \textit{oscillatory bounce}, \textit{matter bounce}, and \textit{exponential bouncing} model settings. Further, we examine the dark energy profiles of the models using reconstructed Lagrangians. In addition, we found that an additional term arises in each reconstructed Lagrangian compared to general relativity (GR). That extra term corrected the background GR to present bouncing cosmology in modified gravity. These newly motivated cosmological models may have an effect on gravitational phenomena at other cosmological scales.\\


\end{abstract}

\maketitle

\date{\today}

\section{Introduction}
A previously thought to be static Universe has now been proven to be expanding at an accelerated rate owing to various observations in  \cite{lss1,lss2,super1,super2,cmb1,cmb2,cmb3,cmb4,cmb5,cmb6,bao,weaklens}. An elementary retrodiction of this fact gives rise to a Universe originating from a point of infinite energy density - the Big Bang, an initial spacetime singularity that requires a knowledge of quantum gravity. The standard Big Bang cosmology, along with this problem of the initial singularity, comes with a number of other problems like the horizon problem, flatness problem, transplanckian problem, etc. The theory of inflation originally proposed by Alan Guth \cite{Alan/1, Alan/2} was exceptionally successful in solving most of the issues but the problem of singularity proved to be persistent. Moreover, Hawking showed that singularities are an essential part of the originally formulated General Relativity \cite{hawking1,hawking2,hawking3}. With a goal to circumvent this singularity, are born Bouncing Cosmologies. This class of theories are completely classical, eliminating the need for a quantum gravitational description, and hence making it possible to discuss the Universe at or even before the $ t=0 $ mark. Instead of retracing the expansion backward in time to an infinitesimal point, the bouncing cosmological scenario assumes that there was a phase at which the Universe was of a minimum (non-zero) size with a maximum (finite) energy beyond which contraction was not possible. In this way, the Universe does not come into existence through a singularity, but with the expansion of a previously contracting phase, hence avoiding the singularity altogether. As in standard Big Bang cosmology, the scale factor and the Hubble parameter are used to study the evolution of a bouncing Universe. The scale factor, in the bouncing scenario, shrinks to a finite value after which it starts to increase. Correspondingly, the Hubble parameter explodes after reaching zero.\\
\indent The bouncing cosmological paradigm has been studied in quite a lot of frameworks, the most appealing of which are those that use scalar fields \cite{scalar1, scalar2}, Loop Quantum Cosmology \cite{lqc1,lqc2,lqc3,lqc4,lqc5} and  modified gravities \cite{frbounce1, frbounce2, ftbounce1, ftbounce2}. Modified gravity is an extension to Einstein's general theory of relativity to account for observed large-scale behavior that is absent in traditional GR. There are predominantly three types of modified gravities -- [i] Curvature based. These include higher powers of the Ricci scalar ($ R $) which is constructed out of the Ricci tensor ($ R_{\mu\nu} $), [ii] Torsion based. These include terms constructed out of the torsion tensor ($ T^{\alpha}_{\beta\gamma}\;=\;\Gamma^\alpha_{\beta\gamma} - \Gamma^\alpha_{\gamma\beta} $) \cite{trinity} and consequently the torsion scalar ($ T $), and [iii] Based on non-metricity or metric non-compatibility. Actions of these include terms constructed out of the non-metricity tensor ($ Q_{\alpha\beta\gamma}\;=\;\nabla_\alpha g_{\beta\gamma} $) and hence the non-metricity scalar ($ Q $).
These theories are hence labeled as $f(R), f(T) \text{and} f(Q)$ respectively. Both $ f(T) \text{and} f(Q) $ theories explain gravity in flat spacetime as a result of torsion and non-metricity respectively as opposed to the curvature picture in conventional GR or $ f(R) $ for that matter. $f(Q)$ gravity, also known as symmetric-teleparallel gravity, is one of the more recent theories of modified gravity. Since the covariant derivative of a tensor tracks the change in that tensor in a particular direction, $ Q_{\alpha\beta\gamma}(=\nabla_\alpha g_{\beta\gamma}) $ measures the change in the metric $ g_{\beta\gamma} $ in some direction $ x^\alpha $. Symmetric teleparallel gravity is a special case of the teleparallel framework such that the metric is non-compatible but torsion vanishes \cite{fq1}. This version of gravity has attracted a lot of interest in recent years after its introduction in \cite{coincident}. Jimenez et al. \cite{cosmofQ} explore the study of cosmological perturbations in the $ f(Q) $ framework. Gadbail et al. \cite{gaurav1} show that any kind of FLRW cosmology can be linked with a specific $ f(Q) $ gravity. It should be noted that while constructing a model, there are various free parameters that need to be fixed with some constraints. This can be achieved by looking at energy condition tests as in \cite{sanjay}. Various studies have been conducted to check its consistency with observational data \cite{fQobs1, fQobs2, fQbbn}. For other interesting works in $ f(Q) $ gravity, check \cite{sanjay2, simran, zinnat,Solanki/2021,fQlss, simranlss,Esposito/2022,Avik}.\\ 
\indent As mentioned above, bouncing cosmology has been studied quite extensively in various scenarios including $ f(R) $ \cite{bambafR, frbounce1, frbounce2} and $ f(T) $ \cite{ftbounce1, ftbounce2} gravities. Other studies done by Cruz-Dombriz et al. \cite{exttele} and Caruana et al. \cite{fTB} investigated the cosmological bouncing solution for various bouncing models in the framework of extended teleparallel gravity. $ f(R,T) $ gravity has also been used in the analysis of bouncing cosmology in \cite{singhfR, zubairfR}. Now we turn our attention to one of the less explored frameworks -- the $f(Q)$ gravity. While conventional GR and its curvature-based extension $ f(R) $ gravity have been quite successful at explaining the cosmological phenomenon, the occurrence of higher order derivatives of the metric in a pseudo-Riemannian manifold renders the former \textbf{highly non-linear}. The framework of $ f(Q) $ gravity, on the other hand, always gives second-order background field equations. Additionally, this extended symmetric teleparallel equivalent of GR has been tested against and successfully verified the various observational constraints like the Cosmic Microwave Background (CMB), Baryon Acoustic Oscillation (BAO), Observational Hubble Dataset (OHD), etc. (see \cite{fQobs1,fQobs2}) and Big Bang Nucleosynthesis (BBN) \cite{fQbbn}. These arguments along with the fact that $ f(Q) $ does not require an arbitrary quantity called $ \Lambda $ to explain the late time acceleration have popularised the theory in recent years. It is for these reasons that we intend to carry out a study of bouncing cosmological models in this framework of modified $ Q $ gravity.
In the following work, with the help of a reconstruction technique, we aim to study and analyze various scenarios of cosmological bounce in the framework of $f(Q)$ gravity. More specifically, we wish to explore five types of bounces -- symmetric bounce, superbounce, oscillatory bounce, matter bounce, and finally the exponential model II. The reconstruction method employed here works by starting with the explicit form of the scale factor corresponding to a specific type of bounce, which is used to calculate the Hubble parameter $ H(t) $ and hence the non-metricity scalar $ Q $. The expression for $ H(t) $ is used to calculate the (overall) equation of state $ \omega_Q  $ following which, a change of variables is performed from $ t\longrightarrow Q $ so that $ a(t) $ becomes $ a(Q) $. The conservation equation and the first Friedmann equation (for $ f(Q) $) are used to obtain a differential equation in $ F(Q) $. We would like to examine the nature of dark energy in all these models. The explicit expression for $ F(Q) $ can be used to calculate the equation of state parameter for dark energy ($ \omega_Q $) for a potential comparison with other successful models like the $ \Lambda $CDM.\\ 
\indent This manuscript is organized as follows. After giving a brief review of $ f(Q) $ gravity in section \ref{section 2}. We turn to the reconstruction of the Lagrangian in the $ f(Q) $ framework that results in various bouncing cosmologies in section \ref{section 3}. The reconstruction process of the Lagrangian for a bouncing cosmological scenario is explored wherein the subsections discuss various models described above for each of which, a form of the scale factor is assumed corresponding to the ``type" of bounce. Subsequently, the Hubble parameter and the equation of state parameter are calculated which prove to be useful in reconstruction. Using the Friedmann equation and the conservation equation, the desired model is obtained corresponding to that scale factor, and the equation of state parameter for dark energy is calculated by fixing the free parameters. Section \ref{section 4} concludes the article by summarizing the important consequences of this work.\\


\section{$f(Q)$ gravity theory}
\label{section 2}
Lorentzian Geometry is the foundation of General Relativity, which is established by selecting a symmetric and metric-compatible link. The next link is the Levi-Civita connection, and because of its characteristics, it only generates non-zero curvature; torsion and non-metricity vanish \cite{Nakahara/2003}. The Weitzenbock connection is used by its teleparallel equivalent (TEGR), which signifies zero nonmetricity and curvature. However, when employing geometrodynamics as the underlying mathematical theory for gravity, a different form of relationship can be made. In fact, the most general link is known as metric-affine, and it is given by
\begin{equation}
    \hat{\Gamma}^{\,\sigma}_{\,\,\,\alpha\beta}=\Gamma^{\,\sigma}_{\,\,\,\alpha\beta}+K^{\,\sigma}_{\,\,\,\alpha\beta}+L^{\,\sigma}_{\,\,\,\alpha\beta}
\end{equation}
where $\Gamma^{\,\sigma}_{\,\,\,\alpha\beta}=\frac{1}{2}g^{\sigma\lambda}\left(\partial_{\alpha}g_{\lambda\beta}+\partial_{\beta}g_{\lambda\alpha}-\partial_{\lambda}g_{\alpha\beta}\right)$ is the Levi-Civita connection, $K^{\,\sigma}_{\,\,\,\alpha\beta}=\frac{1}{2}T^{\,\sigma}_{\,\,\,\alpha\beta}+T^{\,\,\,\,\,\,\,\sigma}_{(\alpha\,\,\,\,\,\,\beta)}$ is the contorsion tensor, and $L^{\,\sigma}_{\,\,\,\alpha\beta}=-\frac{1}{2}g^{\sigma\lambda}\left(Q_{\alpha\lambda\beta}+Q_{\beta\lambda\alpha}-Q_{\lambda\alpha\beta}\right)$ is the disformation tensor.\\
Another equivalent version of GR, known as the symmetric teleparallel equivalent of GR (STEGR), represents a relatively unexplored area. Here, one takes into account vanishing curvature and torsion, and the gravitational interaction is described by the non-metricity tensor. The non-metricity tensor $Q_{\sigma\alpha\beta}$ is defined as 
\begin{equation}
\label{1}
Q_{\sigma\alpha\beta}=\nabla_{\sigma}g_{\alpha\beta},
\end{equation}
and the corresponding traces are 
\begin{equation}
\label{2}
Q_{\sigma}=Q_{\sigma\,\,\,\,\alpha}^{\,\,\,\,\alpha}\, ,\,\,\,\,\,\,\,\,\tilde{Q}_{\sigma}=Q^{\alpha}_{\,\,\,\,\sigma\alpha}\,.
\end{equation}
Moreover, the superpotential tensor $P_{\,\,\alpha\beta}^{\sigma}$ is given by
\begin{equation}
\label{3}
4P_{\,\,\alpha\beta}^{\sigma}=-Q^{\sigma}_{\,\,\,\,\alpha\beta}+2Q^{\,\,\,\,\,\,\sigma}_{(\alpha\,\,\,\,\beta)}-Q^{\sigma}g_{\alpha\beta}-\tilde{Q}^{\sigma}g_{\alpha\beta}-\delta^{\sigma}_{(\alpha}\, Q\,_{\beta)},
\end{equation} 
acquiring the trace of a non-metricity tensor as 
\begin{equation}
\label{4}
Q=-Q_{\sigma\alpha\beta}P^{\sigma\alpha\beta}.
\end{equation}
Consider the general action of $f(Q)$ gravity supplemented with Lagrange multipliers given as \cite{Beltran/2018}
\begin{equation}
\label{5}
S=\int \left[\frac{1}{2}f(Q)+\lambda_{\alpha}^{\,\,\,\beta\mu\nu} R^{\alpha}_{\,\,\,\beta\mu\nu}+\lambda_{\alpha}^{\,\,\,\mu\nu} T^{\alpha}_{\,\,\,\mu\nu}+\mathcal{L}_m\right]\sqrt{-g}\,d^4x,
\end{equation}
where $f(Q)$ is an arbitrary function of trace of non-metricity tensor and $g$ is determinant of the metric tensor $g_{\alpha\beta}$, $\lambda_{\alpha}^{\,\,\,\beta\mu\nu}$ are the Lagrange multipliers, and $\mathcal{L}_m$ is the matter Lagrangian density. The reason for the deliberate choice of the non-metricity scalar is that, with the choice $f(Q) = Q$, we recover the purportedly "symmetric teleparallel equivalent of GR".\\
Before going any further, it will be useful to state upfront that our geometrical framework has a flat and torsion-free connection, requiring that it correspond to a pure coordinate transformation from the trivial connection as described in \cite{coincident}. More specifically, the connection can be parameterized with a collection of functions $\xi^{\alpha}$, as seen below
\begin{equation}
    \hat{\Gamma}^{\,\sigma}_{\,\,\,\alpha\beta}=\frac{\partial x^{\sigma}}{\partial\xi^{\mu}}\partial_{\alpha}\partial_{\beta}\xi^{\mu}
\end{equation}
In the above connection, It must be understood that $\xi^{\alpha}=\xi^{\alpha}(x^{\sigma})$ is an invertible relation and $\frac{\partial x^{\sigma}}{\partial\xi^{\mu}}$ is the inverse of the associated Jacobian. By using a generic coordinate transformation and a general affine connection that equals zero (i.e., $\hat{\Gamma}^{\,\sigma}_{\,\,\,\alpha\beta}=0$), we can always select a coordinate $\xi^{\alpha}=x^{\sigma}$. This coordinate is known as the coincident gauge. As a result, the non-metricity is reduced to $Q_{\sigma\alpha\beta}=\partial_{\sigma}g_{\alpha\beta}$.

Now, the definition of the energy-momentum tensor for matter reads
\begin{equation}
\label{6}
T_{\alpha\beta}\equiv-\frac{2}{\sqrt{-g}}\frac{\delta(\sqrt{-g}\mathcal{L}_m)} {\delta g^{\alpha\beta}}.
\end{equation}

The $f(Q)$ gravitational field equation derived by varying action \eqref{5} with respect to the metric is expressed as
\begin{equation}
\label{7}
\frac{2}{\sqrt{-g}}\nabla_{\sigma}\left(f_{Q}\sqrt{-g}\,P^{\sigma}_{\,\,\alpha\beta}\right)+\frac{1}{2}f\,g_{\alpha\beta}+
f_{Q}\left(P_{\alpha\sigma\lambda}Q_{\beta}^{\,\,\,\sigma\lambda}-2Q_{\sigma\lambda\alpha}P^{\sigma\lambda}_{\,\,\,\,\,\,\beta}\right)=- T_{\alpha\beta},
\end{equation}
where $f_Q=\frac{d f}{d Q}$. Moreover, varying equation \eqref{5} with respect to the connection yields:
\begin{equation}
\label{conn}
    \nabla_{\sigma}\lambda_{\mu}^{\,\,\,\alpha\beta\sigma}+\lambda_{\mu}^{\,\,\,\alpha\beta}=\sqrt{-g}f_Q\,P_{\,\mu}^{\,\,\,\alpha\beta}+H_{\,\mu}^{\,\,\,\alpha\beta}
\end{equation}
where $H_{\,\mu}^{\,\,\,\alpha\beta}=-\frac{1}{2}\frac{\delta\mathcal{L}_m}{\delta\Gamma^{\mu}_{\,\,\,\alpha\beta}}$ is the hypermomentum tensor density.\\
By taking into consideration the antisymmetry property of $\alpha$ and $\beta$ in the Lagrangian multiplier coefficients, Eq.\eqref{conn} can be reduced to
\begin{equation}
    \nabla_{\alpha}\nabla_{\beta} \left(f_{Q}\sqrt{-g}\,P_{\,\mu}^{\,\,\,\alpha\beta}+H_{\,\mu}^{\,\,\,\alpha\beta}\right)=0.
\end{equation}

The connection equation of motion can be easily calculated by noticing that the variation of the
connection with respect to $\xi^{\sigma}$ is equivalent to performing a diffeomorphism so that $\delta_{\xi}\hat{\Gamma}^{\,\sigma}_{\,\,\,\alpha\beta}=-\mathcal{L}_{\xi}\hat{\Gamma}^{\,\sigma}_{\,\,\,\alpha\beta}=-\nabla_{\alpha}\nabla_{\beta}\xi^{\sigma}$, where we have used that the connection is flat and torsion-free \cite{cosmofQ}. Furthermore, in the absence of hypermomentum\footnote{If there is no hypermomentum, this is trivially true. Second, we can assume that the hypermomentum is antisymmetric, in which case our assertion is identically true ($H_{\,\mu}^{\,\,\,\alpha\beta}= 0$). Finally, if $H_{\,\mu}^{\,\,\,\alpha\beta}\neq 0$, we consider the assertion to be the hypermomentum conservation law.} \cite{Beltran/2018}, the connection field equations read as
\begin{equation}
\label{8}
\nabla_{\alpha}\nabla_{\beta} \left(f_{Q}\sqrt{-g}\,P_{\,\mu}^{\,\,\,\alpha\beta}\right)=0.
\end{equation}

For the metric and connection equations, one can notice that $\mathcal{D}_{\alpha}T^{\alpha}_{\,\,\,\,\beta}=0$, where $\mathcal{D}_{\alpha}$ is the metric-covariant derivative. In the most general scenario, with a nontrivial hypermomentum, a relationship between the divergence of the energy-momentum tensor and the hypermomentum would be obtained \cite{Harko/2018}.\\
The energy-momentum tensor $T_{\alpha\beta}$ of the matter, which is assumed to be a perfect fluid, is given by
\begin{equation}
\label{9}
T_{\alpha\beta}=(p+\rho)u_{\alpha}u_{\beta}+pg_{\alpha\beta},
\end{equation}
where $p$ and $\rho$ are the pressure and energy density of a perfect fluid, respectively, and $u_{\alpha}$ is a four-velocity vector. \\
We will now focus on an FLRW line element described as a flat, homogeneous, and isotropic Universe
\begin{equation}
\label{10}
ds^2=-dt^2+a^2(t)(dx^2+dy^2+dz^2),
\end{equation}
for which the trace of the non-metricity tensor reads $Q=6H^2$, where $H=\frac{\dot{a}}{a}$ is the Hubble parameter. Here, $a(t)$ is a cosmic scale factor.\\
Using $f(Q) = Q + F(Q)$, the related field equations are as follows:
 
\begin{equation}
\label{f1}
3 H^2=\rho +\frac{F}{2}-Q\,F_Q,
\end{equation} 
\begin{equation}
\label{f2}
\left(2Q\,F_{QQ}+F_Q+1\right)\dot{H}+\frac{1}{4}\left(Q+2Q\,F_Q-F\right)=-2p.
\end{equation}
Here, dot $(.)$ denotes derivative with respect to $t$, $F_Q=\frac{dF}{dQ}$, and $F_{QQ}=\frac{d^2F}{dQ^2}$. \\
In this scenario, we investigate the possibility that the Universe is filled with dust and radiation fluids, and hence
\begin{equation}
    \rho=\rho_m+\rho_r,\,\,\,\,\,p=p_m+p_r=\frac{1}{3}\rho_r,
\end{equation}
where $\rho_m$ and $\rho_r$ are the energy densities of the dust and radiation, respectively. Then Eqs.\eqref{f1} and \eqref{f2} can be written as
\begin{equation}
    3H^2=\rho_m+\rho_r+\rho_Q,
\end{equation}
\begin{equation}
    2\dot{H}+3H^2=-\frac{\rho_r}{3}-p_Q,
\end{equation}
where $\rho_Q$ and $p_Q$ are the DE density and pressure, respectively, which contributes to the geometry given by
\begin{equation}
    \rho_Q=\frac{F}{2}-Q\,F_Q,
\end{equation}
\begin{equation}
    p_Q=2\dot{H}\left(2Q\,F_{QQ}+F_Q\right)-\rho_Q.
\end{equation}
As a result, the equation of state due to DE is given by 
\begin{equation}
\label{w}
    \omega_Q=\frac{p_Q}{\rho_Q}=-1+\frac{4\dot{H}\left(2Q\,F_{QQ}+F_Q\right)}{F-2Q\,F_Q},
\end{equation}
which shows the phantom $\omega_Q<-1$, and quintessence, $-1<\omega_Q<-\frac{1}{3}$, dominated Universe. Further, we would like to mention here that the $f(Q)$ gravity satisfies the conservation equation
\begin{equation*}
\dot{\rho}+3H(\rho+p)=0.
\end{equation*}
As a result, we investigate various kinds of bouncing solutions in the next sections and reconstruct some classes of Lagrangians using the tools mentioned above.

\section{Reconstruction of bouncing cosmology}
\label{section 3}
In this section, we analyze the feasibility of obtaining adequate gravitational Lagrangians $f(Q)$ capable of reproducing the cosmic evolution described by various cosmological bouncing models. Due to its interest, we shall consider the bouncing models, namely \textit{ symmetric bounce}, \textit{Super bounce}, \textit{Oscillatory bounce}, \textit{Matter bounce}, and \textit{Exponential model II}. This method makes it possible to solve for the gravitational Lagrangian in accordance with a preferred cosmology,  which may be determined either analytically using the form $a(t)$ or $H(t)$ or by cosmological observations. Both approaches, however, have limitations. In most cases, $a(t)$ or $H(t)$ behavior is only relevant or known during specific times. As a result, the reconstructed Lagrangian's applicability is limited because it only signifies its probable approximate form during specific periods \cite{Capozziello/2015}. For a complete picture, the reconstructed Lagrangian must match the behavior over a large number of cosmological epochs, either by combining different observations or by reconstructing the unification of different epochs, as done in Refs. \cite{Nojiri/2006,Nojiri/2007}.

\subsection{Model I: Symmetric Bounce}
The symmetric bounce model was initially studied in ref. \cite{Cai/2012} to construct a non-singular bouncing cosmology following an ekpyrotic contraction phase. However, in order to avoid problems with primordial modes not penetrating the Hubble horizon, this bounce must be paired with other cosmic behaviors \cite{Bamba/2014,Nojiri/2016,Cai/2014}.
First, we took into account the symmetric bouncing cosmology, which is devoid of singularities and in which the scale factor drops to a (non-zero) minimum, therefore completely avoiding a Big Bang-like singularity \cite{Bamba/2014}. The symmetric bouncing cosmology is distinguished by a scale factor with an exponential evolution,

\begin{equation}
\label{11}
a(t)=A \exp\left(\alpha \frac{t^2}{t^2_*}\right),
\end{equation}
where $A$ and $\alpha$ are positive constants, and $t_*$ is some arbitrary time. In this scenario, the Hubble parameter $H$ and non-metricity $Q$ take simple forms
\begin{equation}
\label{12}
H=\frac{2\alpha t}{t^2_*},\,\,\,\,\,\,Q=\frac{24\alpha^2t^2}{t^4_*}.
\end{equation} 

\begin{figure}[H]
\centering
\subcaptionbox{Plot of $a$ versus $t$.}{\includegraphics[width=0.31\textwidth]{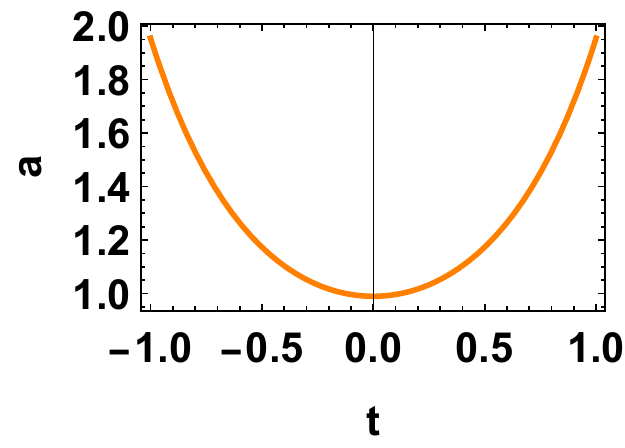}}%
\hfill 
\subcaptionbox{Plot of $H$ versus $t$.}{\includegraphics[width=0.32\textwidth]{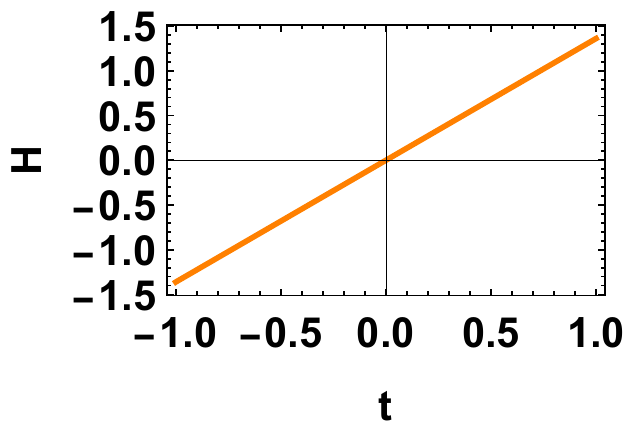}}\hfill 
\subcaptionbox{Plot of $w$ versus $t$.}{\includegraphics[width=0.3\textwidth]{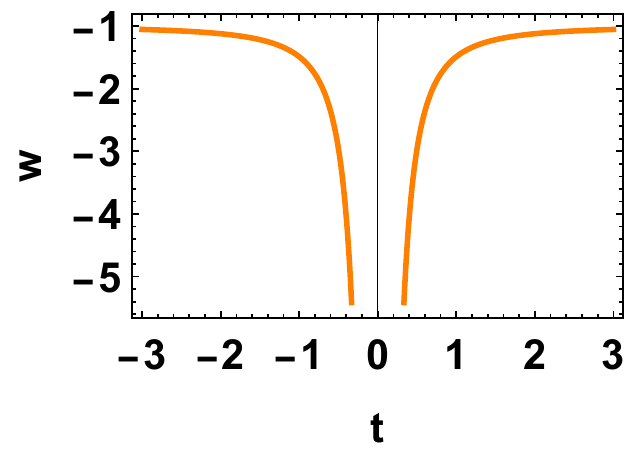}}

\caption{In the above figures, we see that the bounce is located at $t=0$. Since $H=0$ at $t=0$, this value of the Hubble parameter indicates the bouncing point, $H$ is negative at $t<0$ and turns positive at $t > 0$, as seen in fig. (b). In Fig. (c), the EoS parameter is singular at the bouncing point and evolves rapidly near the bounce. In this case, the EoS parameter is symmetric about the bouncing epoch and evolves in the phantom region.}
\end{figure}
Moreover, the scale factor can be expressed in terms
of non-metricity scalar $Q$ as
\begin{equation}
\label{13}
a(Q)=A \exp\left( \frac{Q\, t^2_*}{24\alpha}\right).
\end{equation}

Also, $Q_*=Q(t=t_*)=\frac{24\alpha^2}{t^2_*}$. Then Eq. \eqref{13} is written as
\begin{equation}
\label{14}
a(Q)=A \exp\left( \frac{\alpha\, Q}{Q_*}\right).
\end{equation}
For the EoS $p=w_i\rho$ ($i=$$m$, $r$), the conservation equation lead to
\begin{equation}
\label{15}
\rho=\sum_i \rho_{i0}a^{-3(1+w_i)}.
\end{equation}
Using Eq. \eqref{14}, the above equation can be written as
\begin{equation}
\label{16}
\rho=\frac{Q_0}{2}\sum_i \Omega_{w_i0}A^{-3(1+w_i)}\exp\left(-\frac{3\alpha(1+w_i)Q}{Q_*}\right),
\end{equation}
where $Q_0=6H_0^2$.\\
Putting the above equation in the first Friedman equation of $f(Q)$ gravity, we get

\begin{equation}
\label{17}
\frac{F(Q)}{2}-Q\frac{dF(Q)}{dQ}=-\frac{Q_0}{2}\sum_i \Omega_{w_i0}A^{-3(1+w_i)}
\exp\left(-\frac{3\alpha(1+w_i)Q}{Q_*}\right)+\frac{Q}{2}.
\end{equation}
The solution of the above differential equation is 
\begin{equation}
\label{18}
F(Q)=-Q+c_1\sqrt{Q}-Q_0\sum_i \Omega_{w_i0}A^{-3(1+w_i)}\left[
 \exp\left(-\frac{3\alpha(1+w_i)Q}{Q_*}\right)- \frac{\sqrt{3\pi\, Q_0\,\alpha(1+w_i)\,Q}}{\sqrt{Q_*}}\times Erf\left[\frac{\sqrt{3\alpha(1+w_i)\,Q}}{\sqrt{Q_*}}\right]\right],
\end{equation}
where $Q_0=6H_0^2$, $Erf$ is a error function, and $c_1$ is a integration constant.
\begin{figure}[H]
\centering
\subcaptionbox{Plot of $\omega_Q$ versus $t$ for $\omega_m=0$ and $\Omega_{m0}=0.3$.}{\includegraphics[width=0.4\textwidth]{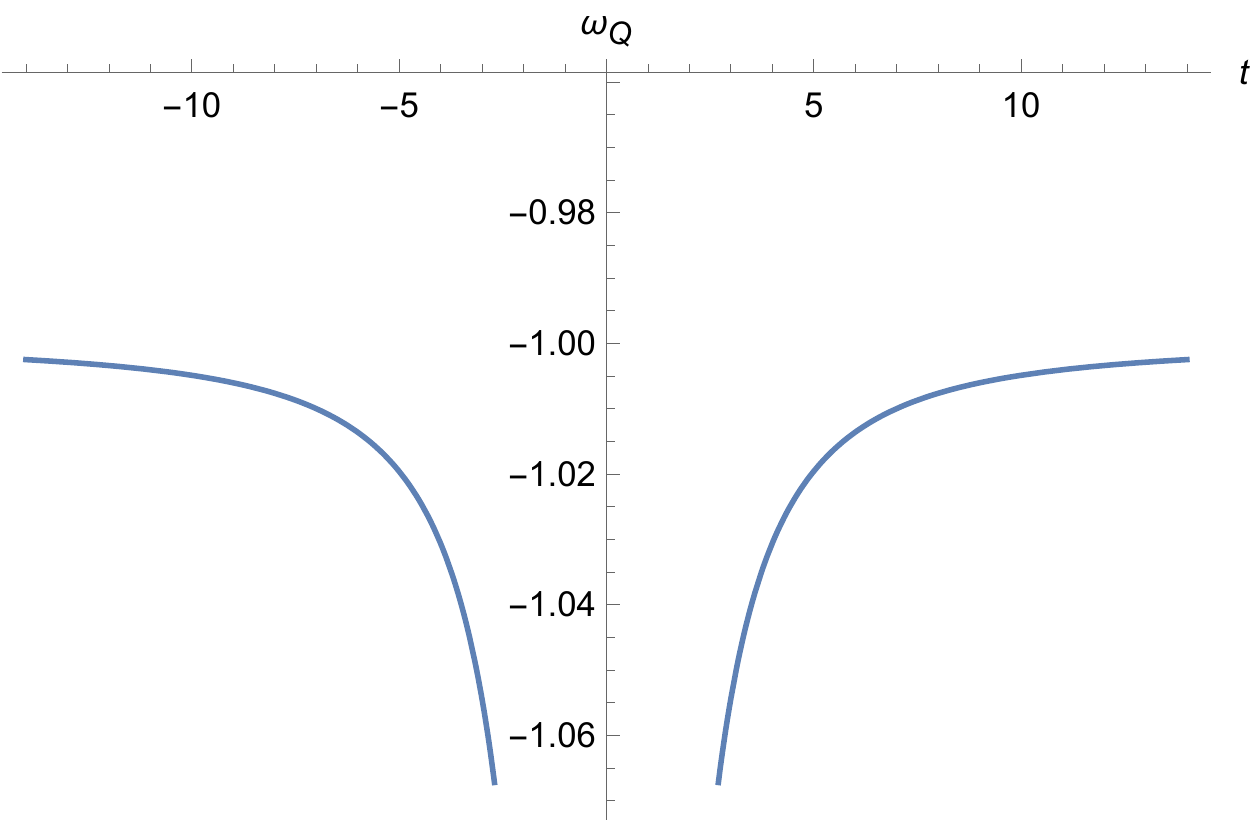}}
 \hspace{0.4in} 
\subcaptionbox{Plot of $\omega_Q$ versus $t$ for $\omega_r=1/3$ and $\Omega_{r0}=0.00005$.}{\includegraphics[width=0.4\textwidth]{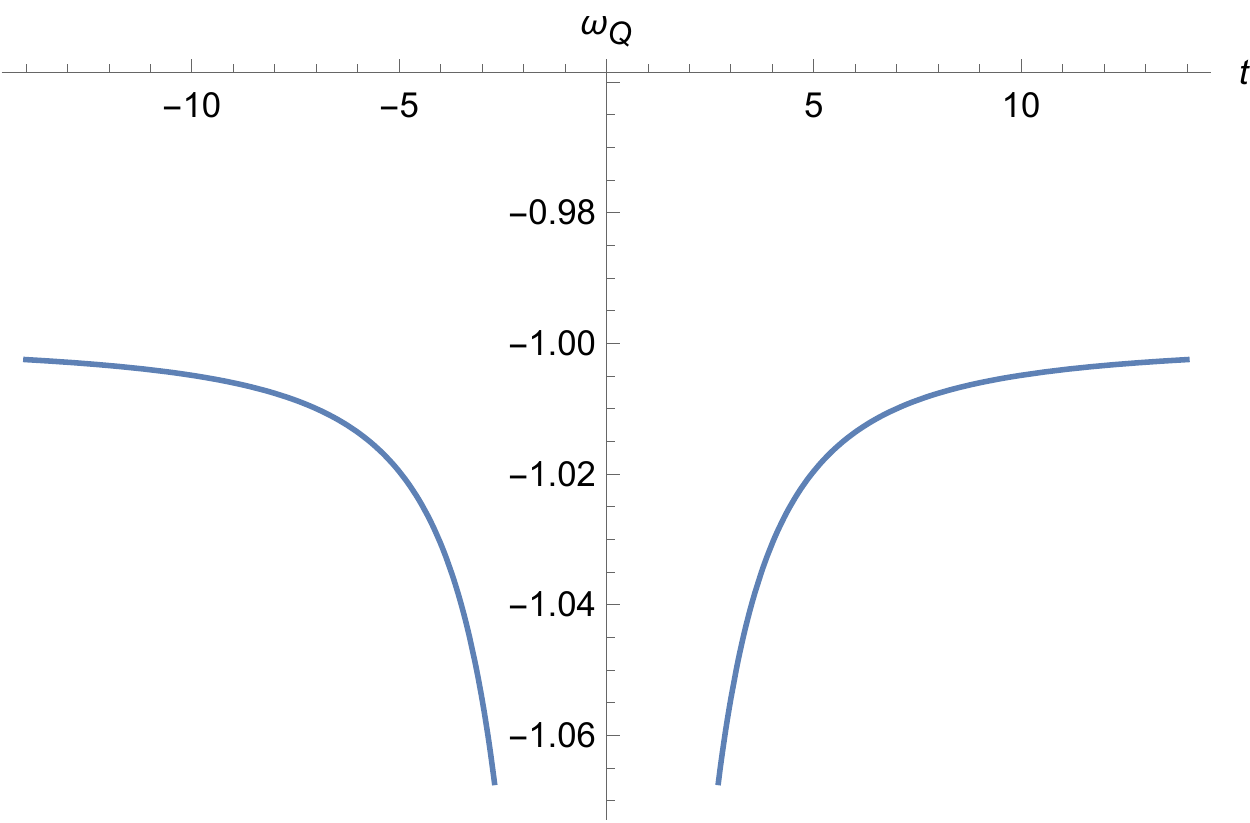}}%
\\ 
\caption{ In Figs. (a) and (b), we can see the evolutionary trajectory of the DE equation state parameter for values $\alpha=0.12$ and $A=1$. }
\label{M1}
\end{figure}

For our reconstructed model \eqref{18}, we check the behavior of the DE equation of state parameter $\omega_Q$ for matter and radiation cases. We can see that in Fig. \ref{M1}, for both cases, the trajectory of the DE equation of state parameter is the same, and it wholly lies in the phantom region ($\omega_Q<-1$) and, at the late-time, it is converging to $-1$. Furthermore, the DE EoS parameter is singular at the bouncing point and rapidly evolves near the bounce. As a result, our reconstructed model \eqref{18} can show the Universe undergoing a phantom DE phase in the symmetric bouncing cosmology.

\subsection{Model II: Superbounce}
A power-law scale factor distinguishes the Super bouncing cosmology, and it is initially considered in \cite{Koehn/2014}. This bouncing model is utilized to build a cosmos that collapses and rebirths without a singularity \cite{Oikonomou/2015}. The Superbounce scale factor is written as 
\begin{equation}
\label{11}
a(t)=\left(\frac{t_s-t}{t_0}\right)^{\frac{2}{c^2}},
\end{equation}
where $c>\sqrt{6}$ is a constant, $t_s$ represents the time at which the bounce occurs and $t_0 > 0$ is an arbitrary time at which the scale factor has a unitary value when $t = t_s +t_0$.
In this case, we have the following expressions for the Hubble parameter

\begin{equation}
H=-\frac{2}{c^2}\left(\frac{1}{t_s-t}\right).
\end{equation}
\begin{figure}[H]
\centering
\subcaptionbox{Plot of $a$ versus $t$.}{\includegraphics[width=0.31\textwidth]{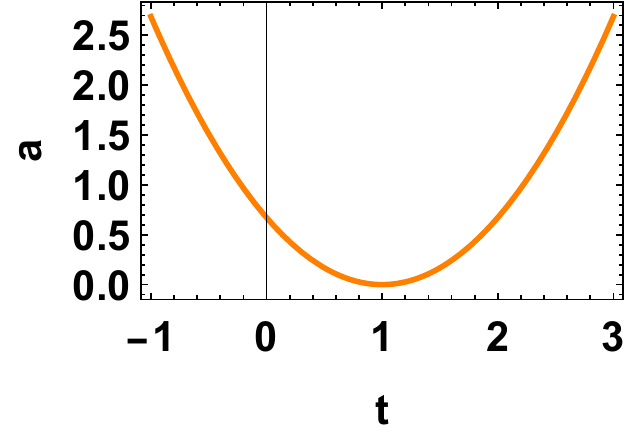}}
 \hspace{0.4in} 
\subcaptionbox{Plot of $H$ versus $t$.}{\includegraphics[width=0.32\textwidth]{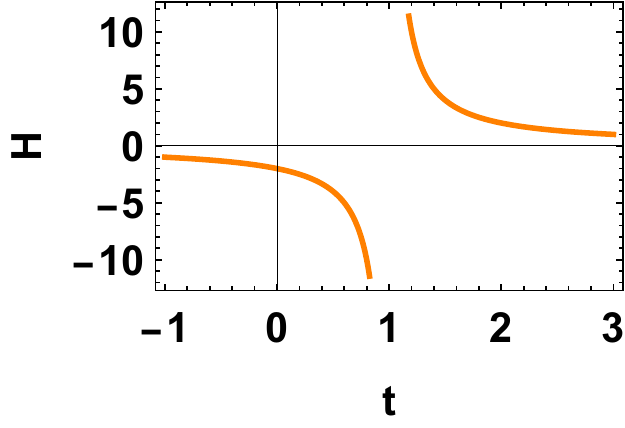}}%
\\ 
\caption{In the above figures, we see that the bounce occurs at $t=t_s$. In Figure (b), we see that Super bounces are characterized by Hubble parameters that change signatures pre- and post-bounce but become singular at the point of bounce. The EoS parameter is constant in the super bounce case, so it is not singular during the bounce. }
\end{figure}
For more simplicity we defined the quantity $t_*=t-t_s$ (leading the bounce to occur at $t_* = 0$) and $\alpha=\frac{2}{c^2}$. Then the expressions for the Hubble parameter and non-metricity scalar are given to be
\begin{equation}
H=\alpha\frac{1}{t_*},\,\,\,\,\,\,Q=\frac{6\alpha^2}{t_*^2}.
\end{equation}
Moreover, the scale factor can be expressed in terms
of non-metricity scalar $Q$ as

\begin{equation}
\label{22}
a(Q)=\left(\frac{6\alpha^2}{t_0^2Q}\right)^{\frac{\alpha}{2}}.
\end{equation}
Also, $Q_0=Q(t_*=t_0)=\frac{6\alpha^2}{t_0^2}$. Then Eq. \eqref{22} is written as 
\begin{equation}
\label{23}
a(Q)=\left(\frac{Q_0}{Q}\right)^{\frac{\alpha}{2}}.
\end{equation}
Using Eq. \eqref{23}, the Eq. \eqref{15} can be written as, 
\begin{equation}
\rho=\frac{Q_0}{2}\sum_i \Omega_{w_i0}\left(\frac{Q_0}{Q}\right)^{\frac{-3\alpha(1+w_i)}{2}}.
\end{equation}
Putting the above equation in the first Friedman equation of $f(Q)$ gravity, we get
\begin{equation}
\frac{F(Q)}{2}-Q\frac{dF(Q)}{dQ}=-\frac{Q_0}{2}\sum_i \Omega_{w_i0}\left(\frac{Q_0}{Q}\right)^{\frac{-3\alpha(1+w_i)}{2}}+\frac{Q}{2}.
\end{equation}
The solution of the above differential equation is 

\begin{equation}
\label{28}
F(Q)=-Q+c_1 \sqrt{Q}-Q_0\sum_i \frac{ \Omega_{w_i0} \left(\frac{Q_0}{Q}\right)^{\frac{-3 \alpha  (w_i+1)}{2}}}{3 \alpha  (w_i+1)-1},
\end{equation}
where $Q_0=6H_0^2$, and $c_1$ is a integration constant.\\
We differentiate the above model \eqref{28} into two cases: the matter ($m$) case and the radiation ($r$) case.
\begin{itemize}
    \item For matter case $\omega_m=0$,
    \begin{equation}
        F(Q)=-Q+c_1 \sqrt{Q}+\frac{Q_0 \,\Omega_{m0}  \left(\frac{Q_0}{Q}\right)^{-\frac{3 \alpha }{2}}}{1-3 \alpha },\,\,\,\,\,\alpha\neq 1/3
    \end{equation}
    \item For radiation case $\omega_r=1/3$,
    \begin{equation}
        F(Q)=-Q+c_1 \sqrt{Q}+\frac{Q_0 \,\Omega_{r0}\left(\frac{Q_0}{Q}\right)^{-2 \alpha }}{1-4 \alpha },\,\,\,\,\,\alpha\neq 1/4
    \end{equation}
\end{itemize}

\begin{figure}[H]
\centering
\subcaptionbox{Plot of $\omega_Q$ versus $t$ for $\omega_m=0$ and $\Omega_{m0}=0.3$.}{\includegraphics[width=0.4\textwidth]{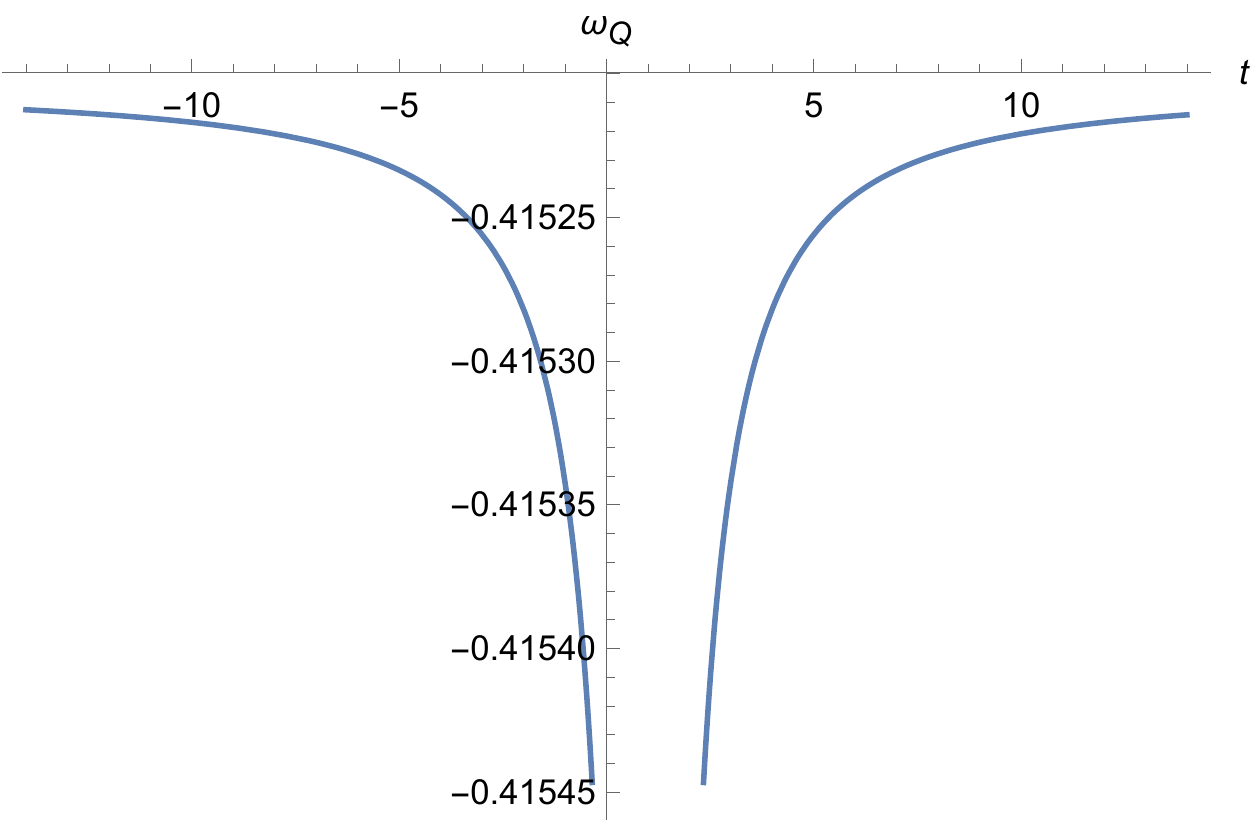}}
 \hspace{0.4in} 
\subcaptionbox{Plot of $\omega_Q$ versus $t$ for $\omega_r=1/3$ and $\Omega_{r0}=0.00005$.}{\includegraphics[width=0.4\textwidth]{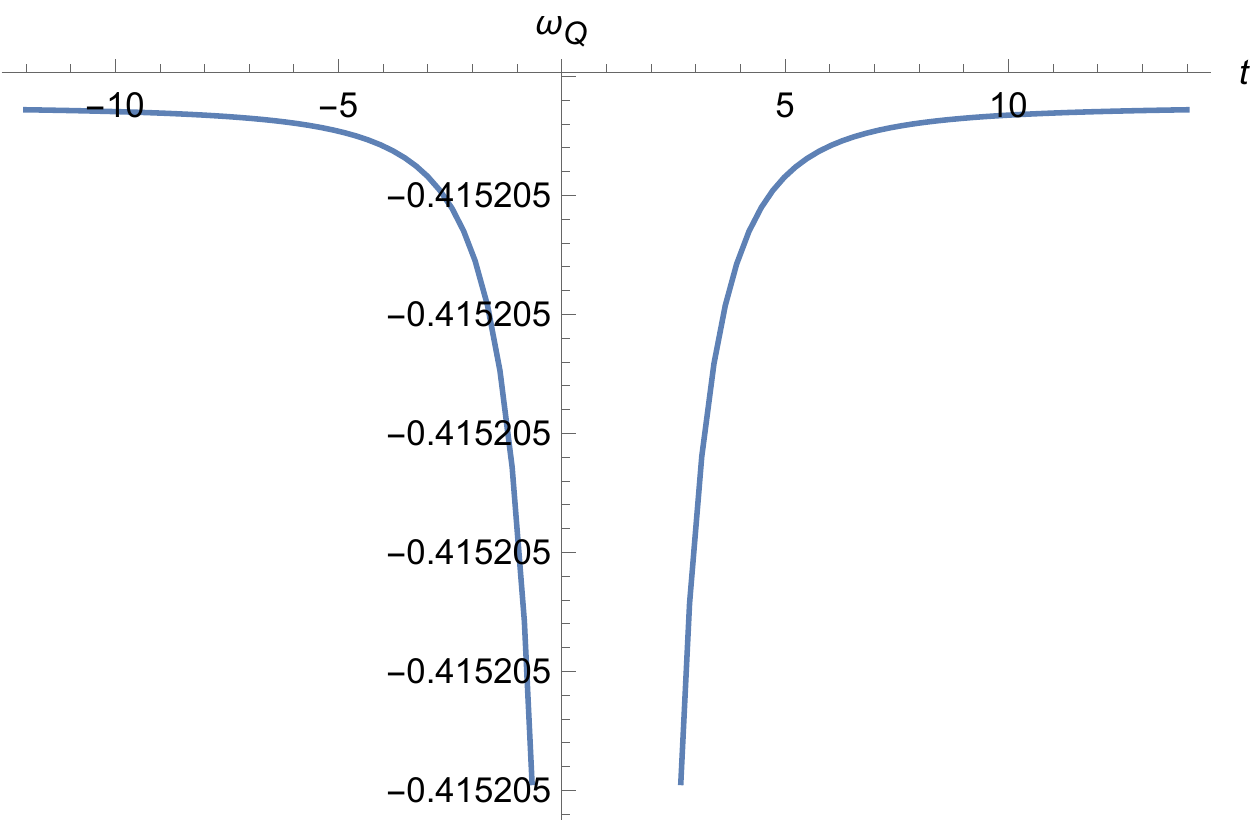}}%
\\ 
\caption{In Figs. (a) and (b), we can see the evolutionary trajectory of the DE equation state parameter for value $\alpha=1.14$. }
\end{figure}
For this symmetric bouncing reconstructed $F(Q)$ model \eqref{28}, the trajectory of the DE equation of state parameter wholly lies in the quintessence region, and it does not cross the phantom divide line $\omega_Q=-1$. At the present time, the value of the DE EoS parameter is $\omega_Q=-0.412$ and $\omega_Q=-0.415$ for matter and radiation cases, respectively, and at the late-time, it is converging to $-0.40$. That means our reconstructed model \eqref{28} can show the Universe undergoing a quintessence DE phase in the super-bouncing cosmology.
 
\subsection{Model III: Oscillatory Bouncing}
According to this model, the Universe expands and contracts periodically. Each cycle begins with a "Big Bang," finishes with a "big crunch," and then begins again with a "Big Bang" \cite{Tolman,Steinhardt/2002,Khoury/2004}.\\ 
The Oscillatory bouncing cosmology is distinguished by a periodic scale factor,
\begin{equation}
\label{11}
a(t)=A \sin^2\left(\frac{B t}{t_*}\right),
\end{equation}
where $t_*>0$ is some reference time, and $A>0$ and $B>0$ are dimensionless constant. In this scenario, the Hubble parameter $H$ and non-metricity $Q$ take simple forms
\begin{equation}
H=\frac{2B}{t_*}\cot\left(\frac{Bt}{t_*}\right),\,\,\,\,\,\,Q=\frac{24B^2}{t_*^2}\cot^2\left(\frac{Bt}{t_*}\right).
\end{equation}
\begin{figure}[H]
\centering
\subcaptionbox{Plot of $a$ versus $t$.}{\includegraphics[width=0.31\textwidth]{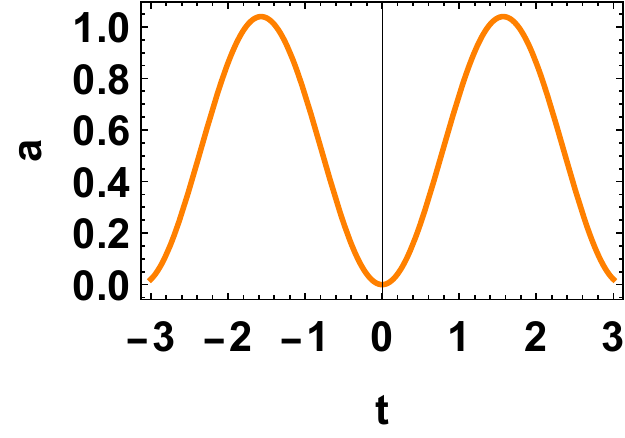}}%
\hfill 
\subcaptionbox{Plot of $H$ versus $t$.}{\includegraphics[width=0.31\textwidth]{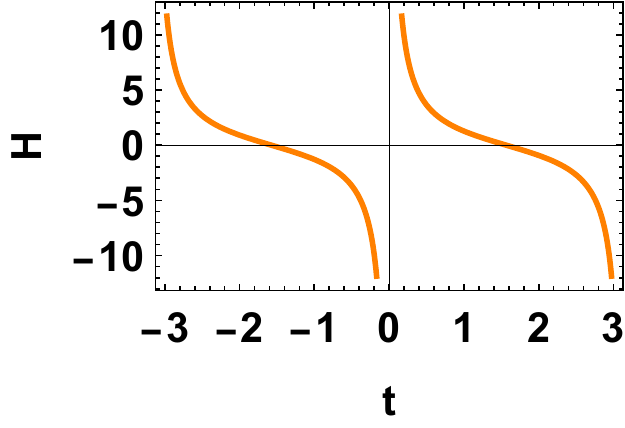}}%
\hfill 
\subcaptionbox{Plot of $w$ versus $t$.}{\includegraphics[width=0.3\textwidth]{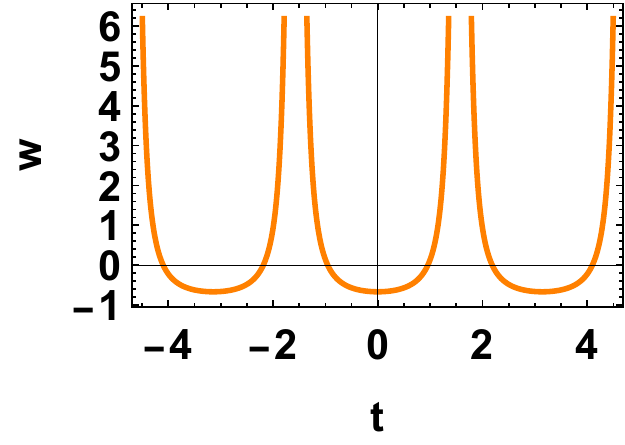}}%
\caption{Figure (a) shows that the oscillatory bounce model generates two different forms of bounce. First occur when $t=\frac{n\pi t_*}{B}$ for $n\in Z$,  corresponds to a Big Crunch/Big Bang singularity. Second occurs when $t=\frac{(2n+1)\pi t_*}{2B}$ for $n\in Z$ (when the Universe reaches its maximum size with no further expansion). We can see in fig. (b), at $t=\frac{n\pi t_*}{C}$ for $n\in Z$, $H$ is singular and at $t=\frac{(2n+1)\pi t_*}{2c}$ for $n\in Z$, $H$ transits from positive to negative. }
\label{osc}
\end{figure}
Moreover, the scale factor can be expressed in terms
of non-metricity scalar $Q$ as
\begin{equation}
\label{29}
a(Q)=\frac{24A\,B^2}{24B^2+Q\,t^2_*}.
\end{equation}

Using Eq. \eqref{29}, the Eq. \eqref{15} can be written as,
\begin{equation}
\rho=\frac{Q_0}{2}\sum_i \Omega_{w_i0}A^{-3(1+w_i)}\left(1+\frac{t_*^2Q}{24 B^2}\right)^{3(1+w_i)}.
\end{equation}

Putting the above equation in the first Friedman equation of $f(Q)$ gravity, we get
\begin{equation}
\frac{F(Q)}{2}-Q\frac{dF(Q)}{dQ}=\frac{Q}{2}
-\frac{Q_0}{2}\sum_i \Omega_{w_i0}A^{-3(1+w_i)}\left(1+\frac{t_*^2Q}{24 B^2}\right)^{3(1+w_i)}.
\end{equation}
The solution of the above differential equation is 

\begin{multline}
\label{a}
 F(Q)=-Q +c_1 \sqrt{Q}+Q_0 \sum_i\Omega_{w_i0}  A^{-3 (w_i+1)}\,
 \left(-\,_2F_1\left[-\frac{1}{2},-3 w_i;\frac{1}{2};-\frac{Q\,t_*^2}{24 B^2}\right]+\frac{t_*^2\,Q}{8B^2} \,_2F_1\left[\frac{1}{2},-3 w_i;\frac{3}{2};-\frac{Q\,t_*^2}{24 B^2}\right]+ \right.\\
 \left.\frac{0.0018\,t_*^4\,Q^2}{B^4} \,_2F_1\left[\frac{3}{2},-3 w_i;\frac{5}{2};-\frac{Q\,t_*^2}{24 B^2}\right]+\frac{t_*^6\,Q^3}{69120\,B^6}\,_2F_1\left[\frac{5}{2},-3 w_i;\frac{7}{2};-\frac{Q\,t_*^2}{24 B^2}\right]\right),
 \end{multline}

where $_2F_1$ is a Hypergeometric function. This Hypergeometric function is one of several special functions represented by the Hypergeometric series, which also contains many other special functions as particular or limiting cases. In our reconstructed $f(Q)$ Lagrangian  \eqref{a}, the Hypergeometric function illustrates the influence of the periodic scale factor. In the context of $f(Q)$ gravity, studying the oscillatory bouncing cosmology might depend on the Hypergeometric series.\\
As shown in Fig. \ref{osc}, the scale factor indicates a cyclic universe in which each cycle is divided by a singularity. However, $\rho_Q$ switches signs in this instance. In the case of $\rho_Q<0$, the expansion slows until the cosmology achieves an equilibrium and then begins to contract. When  $\rho_Q>0$, the contraction is retarded until the spacetime reaches the singularity with no contraction rate but positive acceleration. This fact indicates that the singularity might not be "stable" and, thus, one can utilize this solution to study pre-big bang scenarios in $f(Q)$ gravity.  

\begin{figure}[H]
\centering
\subcaptionbox{Plot of $\omega_Q$ versus $t$ for $\omega_m=0$ and $\Omega_{m0}=0.3$.}{\includegraphics[width=0.4\textwidth]{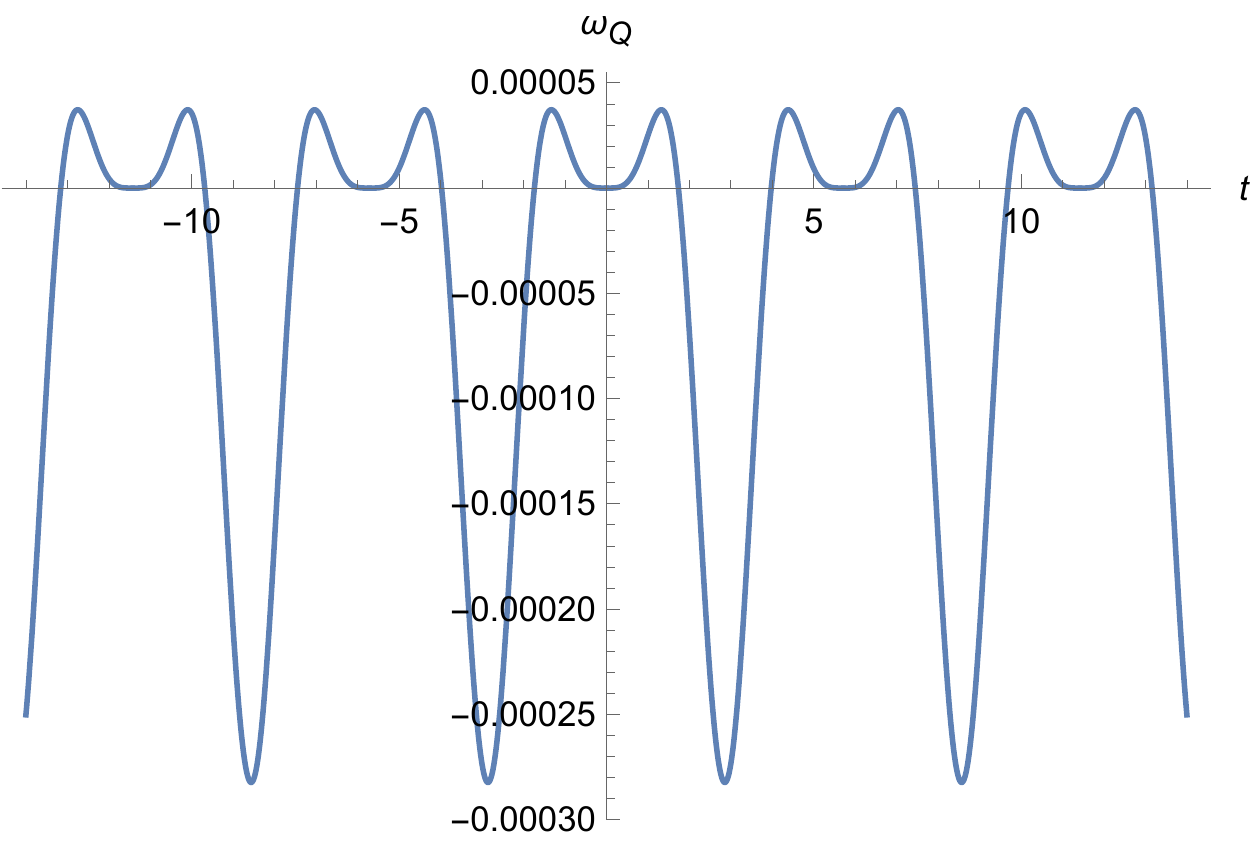}}
 \hspace{0.4in} 
\subcaptionbox{Plot of $\omega_Q$ versus $t$ for $\omega_r=1/3$ and $\Omega_{r0}=0.00005$.}{\includegraphics[width=0.4\textwidth]{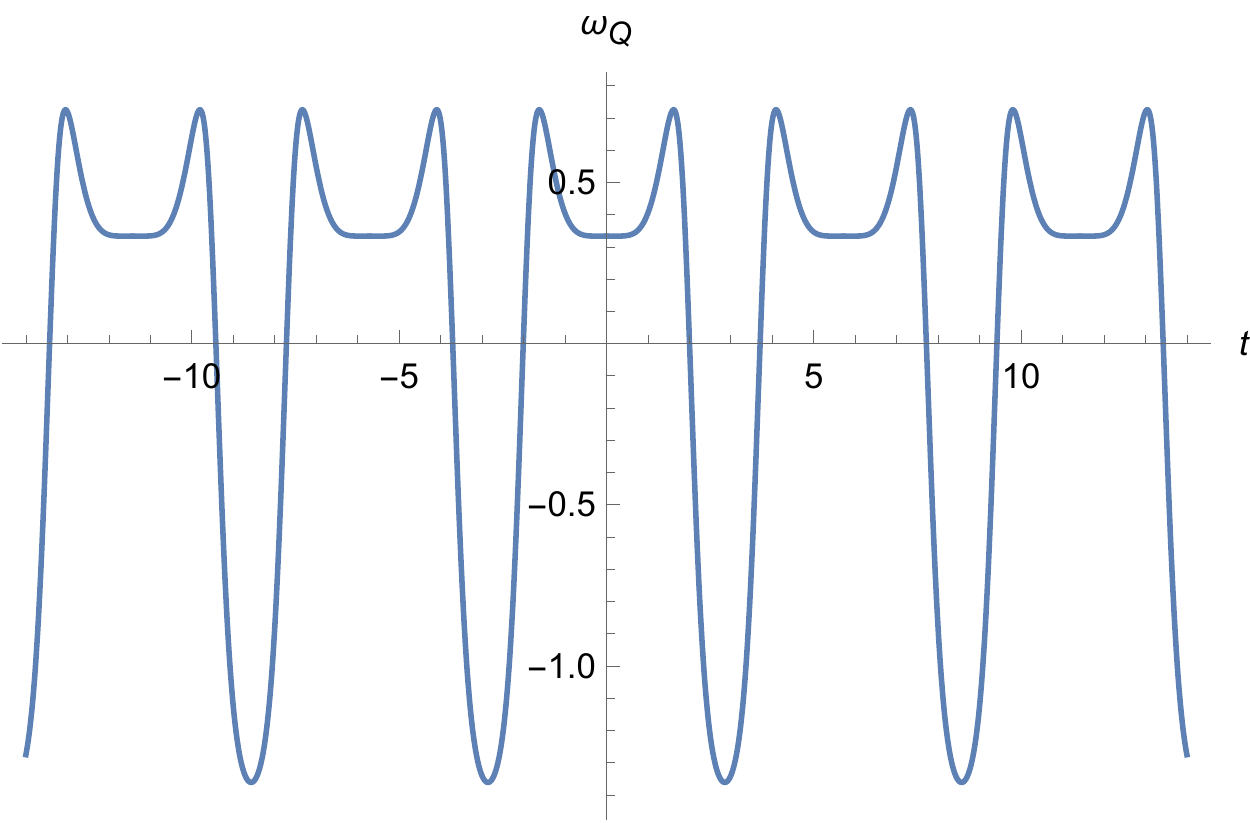}}%
\\ 
\caption{In Figs. (a) and (b), we can see the evolutionary trajectory of the DE equation state parameter for values $A=1$ and $B=0.55$. }
\label{fig6}
\end{figure}
For this oscillatory bouncing reconstructed $F(Q)$ model \eqref{a}, the DE equation of state parameter trajectory is oscillating and taking some finite value at the bouncing point. This implies that our rebuilt model may be able to address the singularity problem. In fig. \ref{fig6}, we can see the rapid contraction and expansion of the universe in which each cycle changes the bouncing point of the universe. In both cases, the present universe is in the expansion phase. For our model, the radiation scenario exhibits better DE EoS parameter behavior than the matter case. At the present time, the value of the DE EoS parameter crosses the phantom divide line ($\omega_Q=-1$) for radiation cases. Moreover, the DE equation of state $\omega_Q$ can show the Universe undergoes a phase transition from $\omega_Q<-1$ to $\omega_Q>-1$ and enters into the hot Big Bang age after the bounce. 

\subsection{Model IV: Matter Bounce}
The next model derives from loop quantum cosmology (LQC) and generates the so-called matter bounce \cite{Singh/2006,Ewing/2013,Matter1,Matter2,Matter3,Matter4}. A plausible alternative to inflation is matter bounce \cite{Brandenberger}, which has the intriguing quality of being compatible with Planck observational evidence \cite{Ade/2014,Ade/2016}. Furthermore, the matter bounce scenario produces an almost scale-invariant primordial power spectrum and results in a matter-dominated period during the late stage of expansion \cite{Cai/2009,Cai/2013,Quintin/2014,Haro/2015}.
For this bouncing model, the scale factor takes the form
\begin{equation}
a(t)=A\left(\frac{3}{2}\rho_ct^2+1\right)^{\frac{1}{3}},
\end{equation}
where $0<\rho_c <<1$ is a critical density and $A$ is a dimensionless constant. Here the Hubble parameter and non-metricity scalar take the following forms,
\begin{equation}
H=\frac{2t\rho_c}{2+3\rho_ct^2},\,\,\,\,\,Q=24\left(\frac{t\rho_c}{2+3\rho_ct^2}\right)^2.
\end{equation}
\begin{figure}[H]
\centering
\subcaptionbox{Plot of $a$ versus $t$.}{\includegraphics[width=0.31\textwidth]{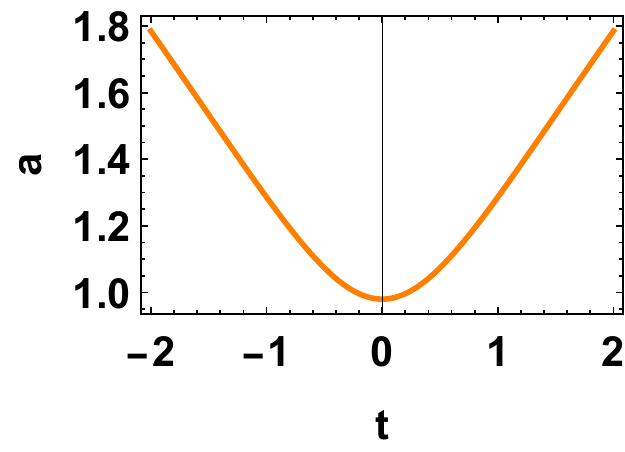}}%
\hfill 
\subcaptionbox{Plot of $H$ versus $t$.}{\includegraphics[width=0.32\textwidth]{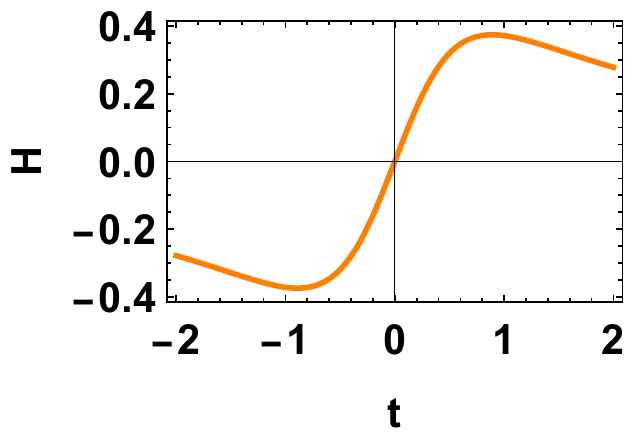}}%
\hfill 
\subcaptionbox{Plot of $w$ versus $t$.}{\includegraphics[width=0.32\textwidth]{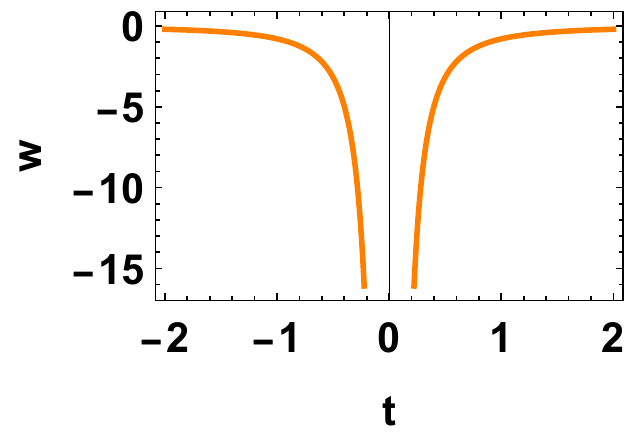}}%
\caption{In figure (b), we see that the Hubble parameter $H < 0$ in the pre-bounce phase, $H = 0$ at the bounce point, and $H > 0$ during the post-bounce epoch. In Fig. (c), the EoS parameter is singular at the bouncing point and evolves rapidly near the bounce. In this case, the EoS parameter is symmetric about the bouncing epoch and evolves in the phantom region.}
\end{figure}
Moreover, the scale factor can be expressed in terms
of non-metricity scalar $Q$ as
\begin{equation}
\label{36}
a(Q)=A\left(\frac{2\rho_c\pm 2\sqrt{\rho_c(\rho_c-Q)}}{Q}\right)^{\frac{1}{3}}.
\end{equation}

Using Eq. \eqref{36}, the Eq. \eqref{15} can be written as,
\begin{equation}
\rho=\frac{Q_0}{2}\sum_i \Omega_{w_i0}A^{-3(1+w_i)}\left(\frac{2\rho_c\pm 2\sqrt{\rho_c(\rho_c-Q)}}{Q}\right)^{-(1+w_i)}.
\end{equation}

Putting the above equation in the first Friedman equation of $f(Q)$ gravity, we get
\begin{equation}
\frac{F(Q)}{2}-Q\frac{dF(Q)}{dQ}=\frac{Q}{2}-\frac{Q_0}{2}\sum_i \Omega_{w_i0}A^{-3(1+w_i)}\times 
\left(\frac{2\rho_c\pm 2\sqrt{\rho_c(\rho_c-Q)}}{Q}\right)^{-(1+w_i)}.
\end{equation}
Here we get two solutions (one for + and second for -) to the above differential equation are 
\begin{widetext}
\begin{multline}
\label{48}
F_+(Q)=-Q+c_1 \sqrt{Q}+\frac{Q_0}{\rho _c}\sum_i  A^{-3 \left(w_i+1\right)} \Omega _{0 w_i} \left(\frac{Q}{\sqrt{\rho _c \left(\rho _c-Q\right)}+\rho _c}\right)^{w_i} \left(2^{-1-\omega_i}\left(-\rho_c+\sqrt{\rho_c(\rho_c-Q)}\right)-\right.\\
\left.\left(\frac{w_i+1}{\sqrt{2}}\right) \left(\rho _c+\sqrt{\rho _c \left(\rho _c-Q\right)}\right)\left(1-\frac{\sqrt{\rho _c \left(\rho _c-Q\right)}}{\rho _c}\right)^{\frac{1}{2}-w_i} \, _2F_1\left[\frac{1}{2},\frac{1}{2}-w_i;\frac{3}{2};\frac{\sqrt{\rho _c \left(\rho _c-Q\right)}+\rho _c}{2 \rho _c}\right]\right),
\end{multline}
and
\begin{multline}
\label{49}
F_-(Q)=-Q+c_1 \sqrt{Q}+\frac{Q_0}{2\rho_c}\sum_iA^{-3\left(w_i+1\right)}\Omega_{w_i0} \left(-2^{-w_i} \left(\frac{\rho_c+\sqrt{\rho_c(\rho_c-Q)}}{\rho_c}\right)^{1+w_i}\right.\\
\left.-\frac{2\sqrt{2}(1+w_i)\,Q}{\sqrt{\rho_c^2+\rho_c\sqrt{\rho_c(\rho_c-Q)}}}\,\,\,_2F_1\left[\frac{1}{2},\frac{1}{2}-w_i;\frac{3}{2};\frac{\rho _c-\sqrt{\rho _c \left(\rho _c-Q\right)}} {2 \rho _c}\right]\right),
\end{multline}
\end{widetext}
where $_2F_1$ is a Hypergeometric function and $c_1$ is a integration constant.
\begin{figure}[H]
\centering
\subcaptionbox{Plot of $\omega_Q$ versus $t$ for $\omega_m=0$ and $\Omega_{m0}=0.3$.}{\includegraphics[width=0.4\textwidth]{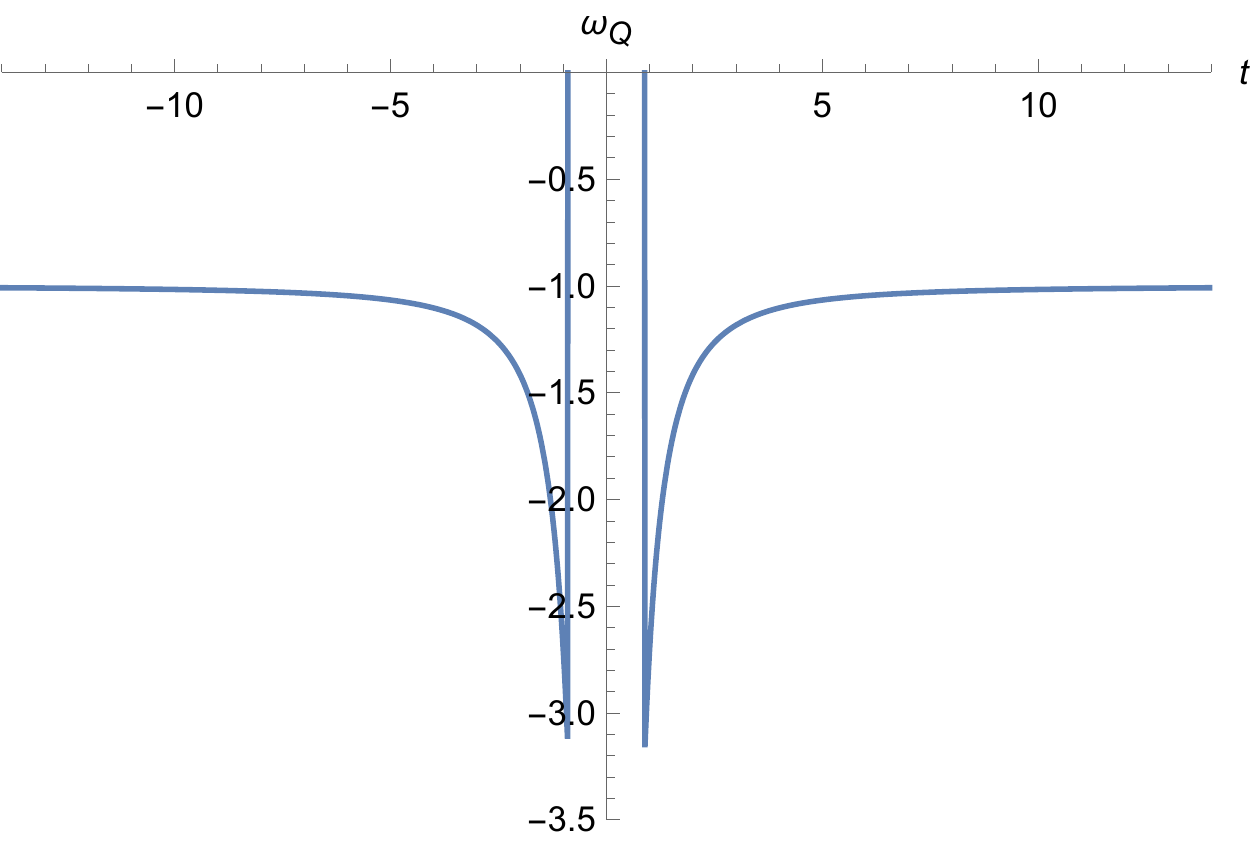}}
 \hspace{0.4in} 
\subcaptionbox{Plot of $\omega_Q$ versus $t$ for $\omega_r=1/3$ and $\Omega_{r0}=0.00005$.} {\includegraphics[width=0.4\textwidth]{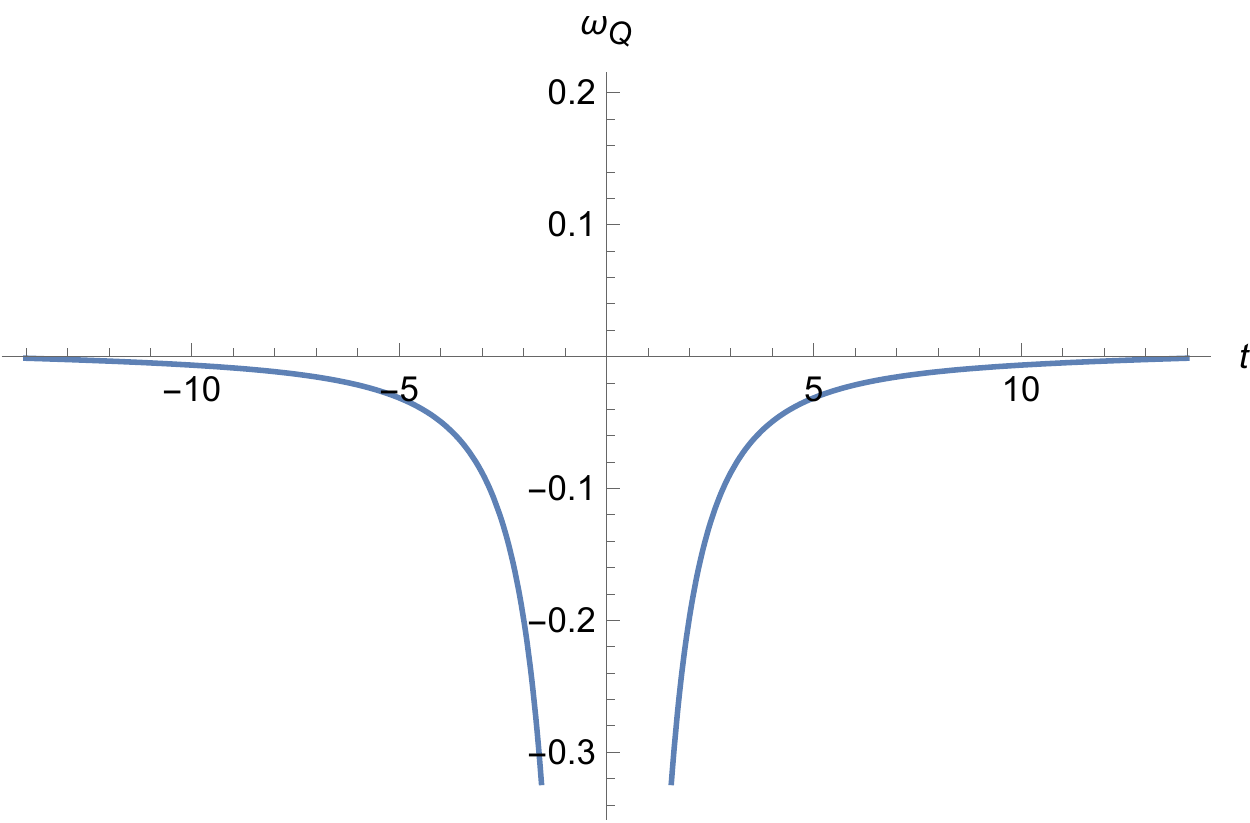}}%
\\ 
\subcaptionbox{Plot of $\omega_Q$ versus $t$ for $\omega_m=0$ and $\Omega_{m0}=0.3$.}{\includegraphics[width=0.4\textwidth]{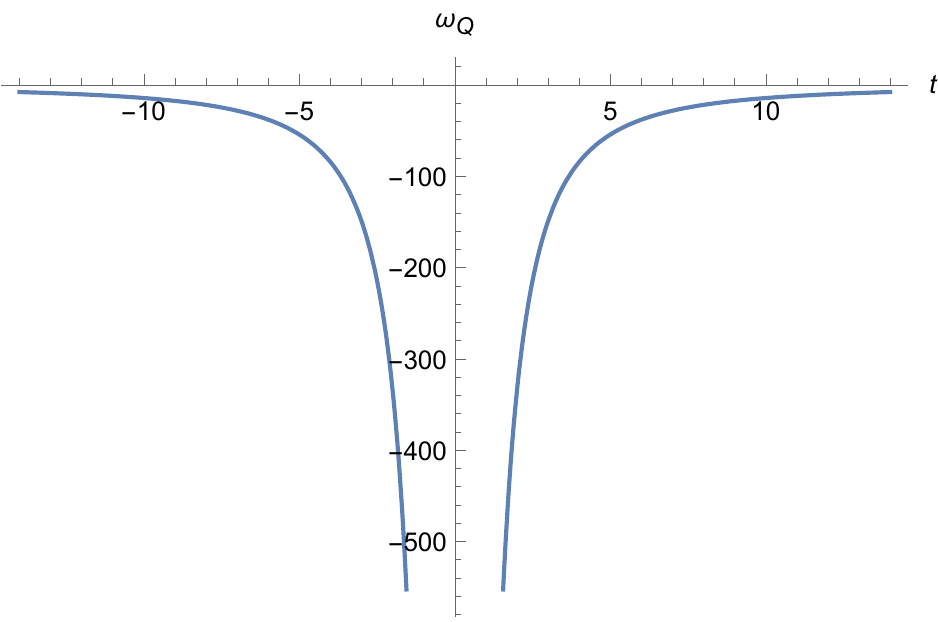}}
 \hspace{0.4in} 
\subcaptionbox{Plot of $\omega_Q$ versus $t$ for $\omega_r=1/3$ and $\Omega_{r0}=0.00005$.} {\includegraphics[width=0.4\textwidth]{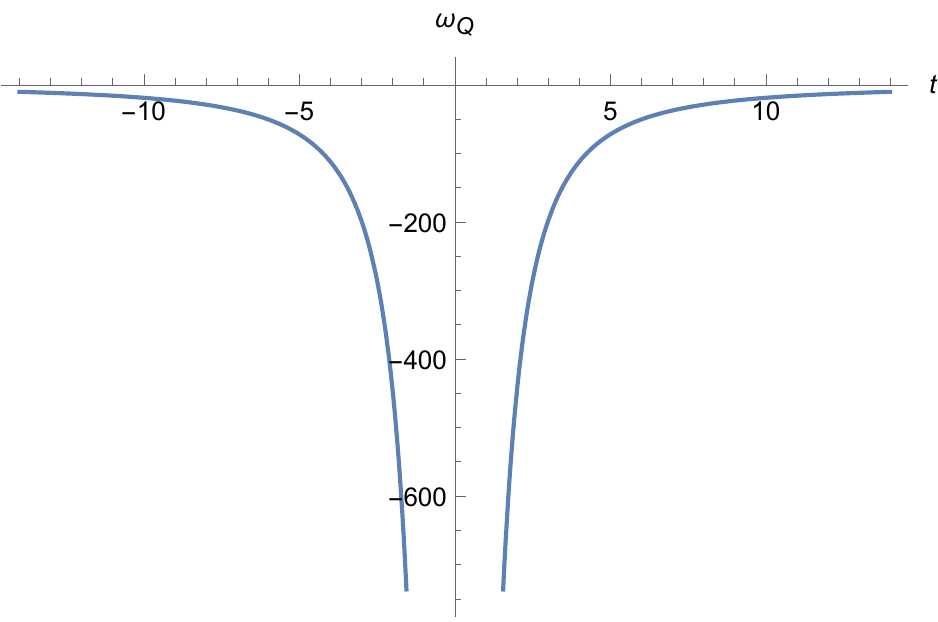}}%
\caption{ Figs. (a) and (b) are obtained for model \eqref{48}, and Figs. (c) and (d) are obtained for model \eqref{49}. In Fig. (a) and (b), we can see the evolutionary trajectory of the DE equation state parameter for values $\rho_c=0.84$ and $A=14$. In Fig. (c) and (d), we can see the evolutionary trajectory of the DE equation state parameter for values $\rho_c=0.0005$ and $A=6$.}
\end{figure}
For corrected model \eqref{48}, the evolutionary trajectory of the DE equation of state $\omega_Q$ lies in the phantom region ($\omega_Q<-1$) for dust case, and at the late-time, it is converging to $-1$. For the radiation case, the evolutionary trajectory of the DE equation of state $\omega_Q$ lies in the quintessence region ($-1<\omega_Q<-1/3$), and at the late-time, it is converging to $-1/3$. As a result, we can say that this model can be viable to study both the quintessence and phantom DE behavior of the Universe.\\
For corrected model \eqref{49}, the evolutionary trajectory of the DE equation of state $\omega_Q$ lies in the phantom region ($\omega_Q<-1$) for both dust and radiation cases. As a result, our corrected model \eqref{49} can study the Universe's phantom DE behavior in matter-bouncing cosmology.

\subsection{Model V: Exponential Model II}
The last bouncing model is comparable to the first one but similar to the power-law model previously discussed, it may also contain a future singularity. For this bouncing model, the scale factor takes the form
\begin{equation}
a(t)=A \exp\left(\frac{f_0}{\alpha+1}(t-t_s)^{(\alpha+1)}\right),
\end{equation}
where $f_0$ and $\alpha$ are some arbitrary constant, $A>0$ is a dimensionless constant and at the bouncing time $t_s$, $a(t_s)=A$. Here the Hubble parameter and non-metricity scalar take the following forms,
\begin{equation}
H=f_0(t-t_s)^{\alpha},\,\,\,\,\,\,Q=6f_0^2(t-t_s)^{2\alpha}.
\end{equation}
\begin{figure}[H]
\centering
\subcaptionbox{Plot of $a$ versus $t$.}{\includegraphics[width=0.31\textwidth]{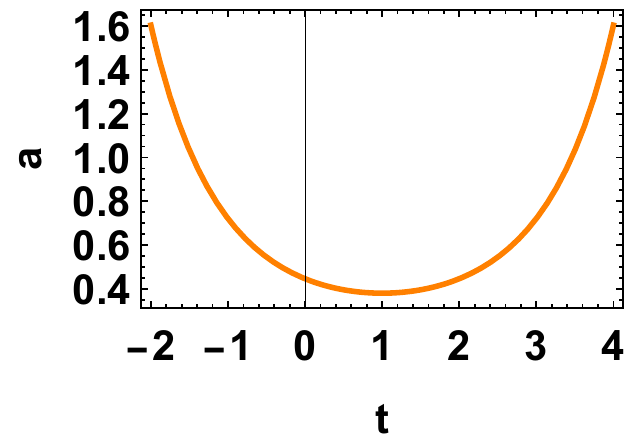}}%
\hfill 
\subcaptionbox{Plot of $H$ versus $t$.}{\includegraphics[width=0.32\textwidth]{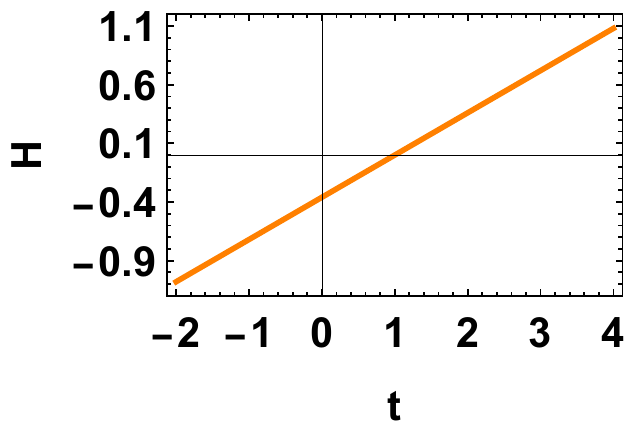}}%
\hfill 
\subcaptionbox{Plot of $w$ versus $t$.}{\includegraphics[width=0.31\textwidth]{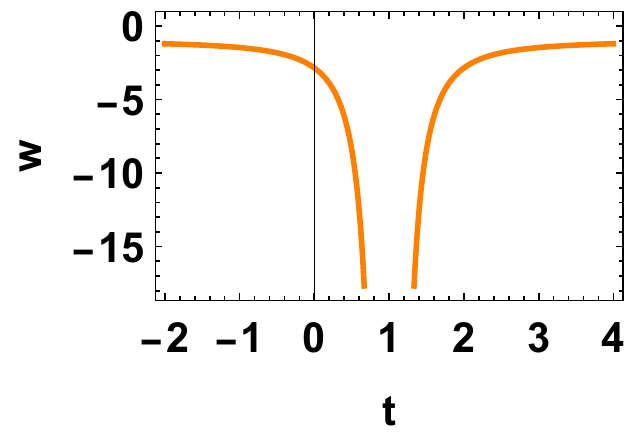}}%
\caption{In the above figures, we see that the bounce occurs at $t=t_s$. Since $H=0$ at the bouncing point, this value of the Hubble parameter indicates the bouncing point, the Hubble parameter $H < 0$ in the pre-bounce phase, and turns positive in the post-bouncing phase, as seen in figure (b). In Fig. (c), the EoS parameter is singular at the bouncing point and evolves rapidly near the bounce. In this case, the EoS parameter is symmetric about the bouncing epoch and evolves in the phantom region.}
\end{figure}
Moreover, the scale factor can be expressed in terms
of non-metricity scalar $Q$ as
\begin{equation}
\label{42}
a(Q)=A \exp\left(\frac{f_0}{\alpha+1}\left(\frac{Q}{6f_0^2}\right)^{\frac{\alpha+1}{2\alpha}}\right).
\end{equation}
In this bouncing cosmology, a type IV singularity (see Ref. \cite{Nojiri/2005}) may occur when
\begin{equation*}
    \alpha=\frac{2n+1}{2m+1},
\end{equation*}
where $m$, $n$ $\in \mathbb{N}$ and $\alpha>1$.\\
Using Eq. \eqref{42}, the Eq. \eqref{15} can be written as,
\begin{equation}
\rho=\frac{Q_0}{2}\sum_i \Omega_{w_i0}A^{-3(1+w_i)}\exp\left(\frac{-3(1+w_i)f_0}{\alpha+1}\left(\frac{Q}{6f_0^2}\right)^{\frac{\alpha+1}{2\alpha}}\right).
\end{equation}

Putting the above equation in the first Friedman equation of $f(Q)$ gravity, we get
\begin{equation}
\frac{F(Q)}{2}-Q\frac{dF(Q)}{dQ}=-\frac{Q_0}{2}\sum_i \Omega_{w_i0}A^{-3(1+w_i)}\times 
\exp\left(\frac{-3(1+w_i)f_0}{\alpha+1}\left(\frac{Q}{6f_0^2}\right)^{\frac{\alpha+1}{2\alpha}}\right)+\frac{Q}{2}.
\end{equation}

The solution of the above differential equation is 

\begin{multline}
\label{55}
F(Q)=-Q+c_1 \sqrt{Q}-\frac{\alpha  Q_0}{\alpha+1}\sum_i \Omega _{ w_i0} A^{-3 \left(w_i+1\right)} \left( \frac{3^{\frac{\alpha -1}{2 \alpha }} \alpha  f_0 \left(w_i+1\right) \left(\frac{Q}{2f_0^2}\right)^{\frac{\alpha +1}{2 \alpha }}}{\alpha+1}\right)^{\frac{\alpha}{\alpha+1}}\Gamma\left[-\frac{\alpha}{\alpha+1}, \frac{3^{\frac{\alpha -1}{2 \alpha }} \alpha  f_0 \left(w_i+1\right) \left(\frac{Q}{2f_0^2}\right)^{\frac{\alpha +1}{2 \alpha }}}{\alpha+1}\right]
\end{multline}

where $\Gamma$ is a Gamma function and $c_1$ is an integration constant.

\begin{figure}[H]
\centering
\subcaptionbox{Plot of $\omega_Q$ versus $t$ for $\omega_m=0$ and $\Omega_{m0}=0.3$.}{\includegraphics[width=0.4\textwidth]{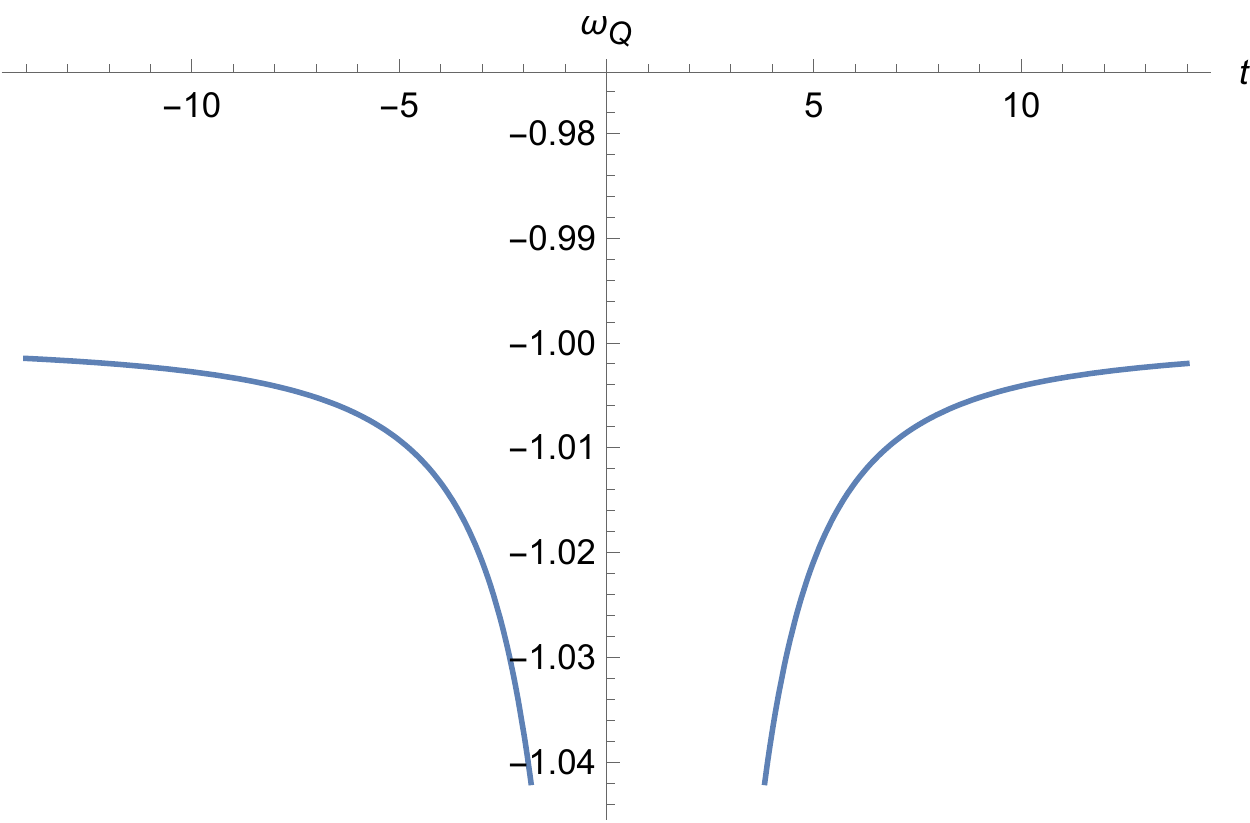}}
 \hspace{0.4in} 
\subcaptionbox{Plot of $\omega_Q$ versus $t$ for $\omega_r=1/3$ and $\Omega_{r0}=0.00005$.}{\includegraphics[width=0.4\textwidth]{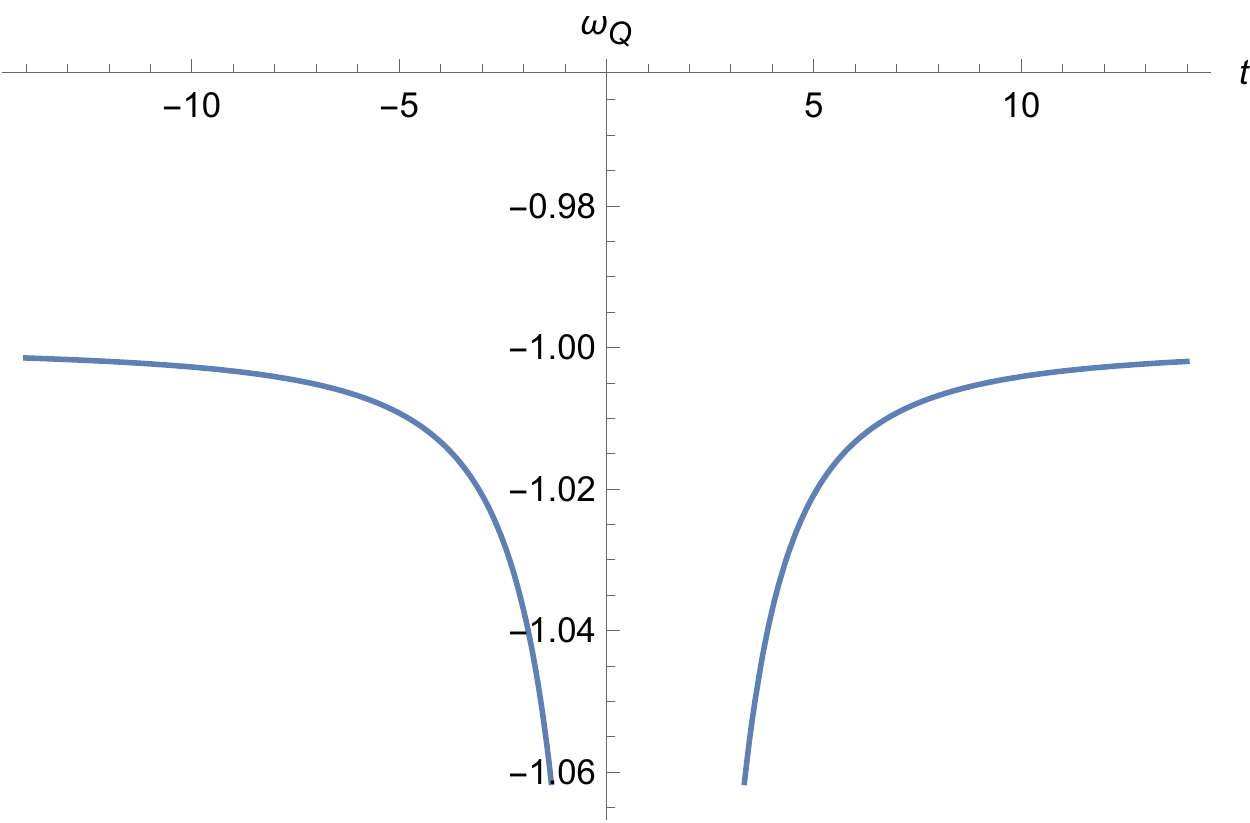}}%
\\ 
\caption{In Figs. (a) and (b), we can see the evolutionary trajectory of the DE equation state parameter for values $\alpha=1$ and $f_0=2$. }
\label{M5}
\end{figure}

 For our reconstructed model \eqref{55},  the behavior of the DE equation of state parameter $\omega_Q$ for matter and radiation cases is the same, and it wholly lies in the phantom region ($\omega_Q<-1$) and, at the late-time, it is converging to $-1$.  In Fig. \ref{M5}, the DE EoS parameter is singular at the bouncing point and evolves rapidly near the bounce. As a result, our reconstructed model \eqref{55} can study the Universe's phantom DE behavior. In the framework of $f(Q)$ gravity, this exponential model II behaves identically to the symmetric bouncing model (or exponential model I) and might be it can give the same types of results but it may include a future singularity.

\section{Conclusion}
\label{section 4}

The initial singularity problem and the inflationary paradigm are widely discussed problems in Big-Bang cosmology. Various approaches have been adopted in the literature to resolve this issue, but bouncing cosmology is one of the best alternative ways to deal with this problem. On the other side, modified gravity is an ideal framework for producing new cosmological models, which can alleviate or entirely eliminate long-lasting cosmological problems. Therefore, in this work, we have aimed to explore this singularity issue within the framework of modified symmetric teleparallel gravity through bouncing cosmology.

Our approach has been to reconstruct the Lagrangians $f(Q)$ against well-known bouncing cosmological solutions in a flat FLRW spacetime with perfect fluid matter distribution. We have explored five different types of bouncing solutions, such as symmetric bounce, super-bounce, oscillatory bounce, matter bounce, and exponential bouncing model, and the profiles of their cosmological parameters. We have not only reconstructed the Lagrangian $f(Q)$ but also used it to check how this modification in the Lagrangian describes the dark energy behavior of the models.

Moreover, it is observed that the scale factor for symmetric bounce, matter bounce, and exponential bouncing model II converge to a finite value, whereas for super-bounce and oscillatory bounce, it converges to zero at the bouncing point. Similarly, we have seen that the expansion rate reduces to zero at the bouncing point in the case of symmetric bounce, matter bounce, and exponential bouncing model, and for the other two models it diverges to infinity. Furthermore, we have tested the dark energy type profiles of the equation of state obtained due to the modifications in the Lagrangian. In particular, we have checked the profiles of the equation of state parameters with the varying leading parameter of each model, such as $\alpha, B, \rho_c, A, f_0$. It is observed that with the positive real numbers of leading parameters, we are getting desired results for all the models except for the oscillatory model, the leading parameter $B$ taking $-2<B<2$ to present a good result. Those profiles helped us explore the effect of modified theory at the bouncing point and the present evolution process of the Universe. It is noted here that $\omega_Q$ shows a phantom type profile near the bouncing point and later converges to $\Lambda$CDM for the model I, model IV (except $f_+$ for radiation), and model V. But $\omega_Q$ lies in the quintessence phase throughout its evolution for model II. We see that the EoS is singular at the bouncing point in all four aforementioned situations (excluding Oscillatory bounce), which appears to be a recurrent pattern, frequent in certain sorts of bouncing cosmologies. We should emphasize that a unique EoS in astrophysical systems is a relatively unusual feature that never occurs, therefore in theory, the same would be predicted in cosmological theories. So we need to figure out what that singularity in the EoS means and how it relates to the finite temporal singularities. Also, in the case of symmetric bounce, matter bounce, and Exponential Model II bounce, the Hubble rate vanishes at the bouncing point, implying that a singularity in the EoS is necessary unless this is somehow neutralized if $H^2$ behaves similarly to $\dot{H}$. Furthermore, note that the effective pressure and effective energy density are not necessarily singular just because the EoS parameter is singular \cite{EoS}. Although they both might be regular. As a result, we could argue that the EoS singularity is not directly related to finite time cosmic singularities \cite{Odintsov/2015}.


It is well known that for Lagrangian $f(Q)=Q+\Lambda \sqrt{Q}$, we cannot be distinguished $f(Q)$ gravity from the GR in the cosmological background evolution \cite{coincident}. In this study, one can see that the reconstructed Lagrangians have one extra term that arises compared to the previous $f(Q)$ function for each case. That additional term helps us to present a corrected expression for $f(Q)$ and might help in avoiding the Big Bang singularity. The advantage of following this method, we do not need to choose any random expression for Lagrangian, which the researcher generally follows to study bouncing cosmology. Further, one may follow a similar procedure to explore other cosmological scenarios. Also, the reconstructed Lagrangians may be used to study other scenarios of the Universe except bouncing cosmology, which might help in constructing singular-free cosmological models. 

In contrast, this study of bouncing cosmologies in the context of modified symmetric teleparallel gravity may shed some light on future work on the early evolution process of the Universe. Here, we have examined the cosmological models that may arise for various bouncing solutions at the background level of cosmology. Further, one may use these reconstructed functions to explore the present cosmological scenario. To be more precise and to present physically relevant models, one may need to study the perturbation for the early Universe and to investigate their impact on the cosmic microwave background. With the reconstructed $f(Q)$ functions, one can find the modifications in the perturbation field equations according to the scalar, vector, and tensor perturbation. Then one can choose a gauge as per the problem considered i.e., the evolution of CMB, neutrinos fluctuations, inflationary cosmology, etc. To study the CMB power spectrum in $f(Q)$ gravity, one may follow some of the techniques presented in \cite{cmbp}. This would find some interesting features of modified symmetric teleparallel gravity. In the near future, we could address some of these studies in the background of this gravity.

\section*{Data Availability Statement}
There are no new data associated with this article.

\section*{Acknowledgments}
GNG acknowledges University Grants Commission (UGC), New Delhi, India for awarding Junior Research Fellowship (UGC-Ref. No.: 201610122060). PKS acknowledges Science and Engineering Research Board, Department of Science and Technology, Government of India for financial support to carry out Research project No.: CRG/2022/001847, IUCAA, Pune, India for providing support through the visiting Associateship program and Transilvania University of Brasov for Transilvania Fellowship for Visiting Professors. We are very much grateful to the honorable referee and to the editor for the illuminating suggestions that have significantly improved our work in terms of research quality, and presentation.\\


\begin{thebibliography}{90}


\bibitem{lss1} M. Tegmark et al., Phys. Rev. D \textbf{69}, 103501 (2004).

\bibitem{lss2} U. Seljak et al., 	Phys. Rev. D \textbf{71}, 103515 (2005).

\bibitem{super1} S. Perlmutter et al., Astrophys. J. \textbf{517}, 565--586 (1999).

\bibitem{super2} Adam G. Riess et al., Astron. J. \textbf{116}, 1009--1038 (1998).

\bibitem{cmb1} D. N. Spergel et al., Astrophys. J. Suppl. \textbf{148}, 175--194 (2003).

\bibitem{cmb2} D. N. Spergel et al., Astrophys. J. Suppl. \textbf{170}, 377 (2007)

\bibitem{cmb3} E. Komatsu et al., 	Astrophys. J. Suppl. \textbf{180}, 330--376 (2009).

\bibitem{cmb4} E. Komatsu et al., Astrophys. J. Suppl. \textbf{192}, 18 (2011).

\bibitem{cmb5} G. Hinshaw et al., Astrophys. J. Suppl. \textbf{208}, 19 (2013).

\bibitem{cmb6} P. A. R. Ade et al., Astron. Astrophys. \textbf{571}, A16 (2014).

\bibitem{bao} D. J. Eisenstein et al., Astrophys. J. \textbf{633}, 560--574 (2005).

\bibitem{weaklens} B. Jain and A. Taylor, Phys. Rev. Lett. \textbf{91}, 141302 (2003).

\bibitem{Alan/1} Alan H. Guth, Phys. Rev. D \textbf{23}(2), 347--356 (1981). 

\bibitem{Alan/2} Alan H. Guth and Paul J. Steinhardt, Sci. Am. \textbf{250}, 116--129 (1984).

\bibitem{hawking1} S. Hawking, Proc. Roy. Soc. Lond. A \textbf{294}, 511--521 (1966).

\bibitem{hawking2} S. Hawking, Proc. Roy. Soc. Lond. A \textbf{295}, 490--493 (1966).

\bibitem{hawking3} S. Hawking, Proc. Roy. Soc. Lond. A \textbf{300}, 187--201 (1967).

\bibitem{scalar1} Y. F. Cai et al.,  J. High Ener. Phys. \textbf{10}, 071 (2007)

\bibitem{scalar2} Y. F. Cai et al., J. Cosmol. Astropart. Phys. \textbf{08}, 020 (2012)

\bibitem{lqc1} A. Ashtekar and P. Singh, Class. Quant. Grav. \textbf{28}, 213001 (2011).

\bibitem{lqc2} M. Bojowald, Class. Quant. Gravit. \textbf{26}, 075020 (2009).

\bibitem{lqc3} T. Cailleteau et al., Phys . Rev. D \textbf{86}, 087301 (2012).

\bibitem{lqc4} A. Corichi and P. Singh, Phys. Rev. D \textbf{80}, 044024 (2009).

\bibitem{lqc5} A. Ashtekar et al., Phys. Rev. D \textbf{74}, 084003 (2006).

\bibitem{frbounce1} M. Bouhmadi-Lopez et al., Phys. Rev. D \textbf{87}, 103528 (2013).

\bibitem{frbounce2} G. Leon and A.A. Roque, J. Cosmol. Astropart. Phys. \textbf{05}, 32 (2014).

\bibitem{ftbounce1} Yi-Fu Cai et ql., Class. Quant. Grav. \textbf{28}, 215011 (2011).

\bibitem{ftbounce2} K. Bamba et al., Phys. Rev. D \textbf{94}, 083513 (2016).

\bibitem{trinity} J. B. Jimenez et al., Universe \textbf{5(7)}, 173 (2019).

\bibitem{fq1} J. M. Nester, H.-J. Yo, Chin. J. Phys. \textbf{37}, 113 (1999).

\bibitem{coincident} J. Beltran Jimenez et al., Phys. Rev. D, \textbf{98}, 044048 (2018).

\bibitem{cosmofQ} J. Beltran Jimenez et al., Phys. Rev. D \textbf{101}, 103507 (2020).

\bibitem{gaurav1} G. N. Gadbail et al., Phys. Lett. B \textbf{835}, 137509 (2022).

\bibitem{sanjay} S. Mandal et al., Phys. Rev. D \textbf{102}, 024057 (2020).

\bibitem{fQobs1} R. Lazkoz et al., Phys. Rev. D \textbf{100}, 104027 (2019).

\bibitem{fQobs2} I. Ayuso et al., Phys. Rev. D \textbf{103}, 063505 (2021).

\bibitem{fQbbn} F. K. Anagnostopoulos, et al., Eur. Phys. J. C \textbf{83}, 58 (2023).

\bibitem{sanjay2} S. Mandal et al., Phys. Rev. D \textbf{102}, 124029 (2020).

\bibitem{simran} S. Arora, P. K. Sahoo, Ann. der Phys. \textbf{534}, 2200233 (2022).

\bibitem{zinnat} Z. Hassan et al., Fort. der Phys. \textbf{69}, 2100023 (2021).

\bibitem{Solanki/2021} R. Solanki et al., Phys. Dark Univ. \textbf{32}, 100820 (2021).

\bibitem{fQlss} I. S. Albuquerque, N. Frusciante, Phys. Dark Univ. \textbf{35}, 100980 (2022). 

\bibitem{simranlss} O. Sokoliuk et al., Mon. Not. Roy. Astron. Soc. \textbf{522}, 252--267 (2023).

\bibitem{Esposito/2022} F. Esposito et al., Phys. Rev. D \textbf{105}, 084061 (2022).

\bibitem{Avik} Avik De et al.,  Eur. Phys. J. C \textbf{82}, 72 (2022).

\bibitem{bambafR} K. Bamba et al., J. Cosmol. Astropart. Phys. \textbf{01}, 008 (2014).


\bibitem{exttele} A. de la Cruz-Dombriz et al., Phys. Rev. D \textbf{97}, 104040 (2018).

\bibitem{fTB} Maria Caruana et al., Eur. Phys. J. C \textbf{80}, 640 (2020).

\bibitem{singhfR} J. K. Singh et al., Phys. Rev. D \textbf{97}, 123536 (2018).

\bibitem{zubairfR} M. Zubair et al., Int. J. Mod. Phys. D \textbf{31}, No. 12, 2250092 (2022).

\bibitem{Nakahara/2003} M. Nakahara, Geometry, Topology and Physics (Taylor and Francis, 2003).

\bibitem{Beltran/2018} J. Beltran Jimenez, L. Heisenberg, and T. S. Koivisto, J. Cosmol. Astropart. Phys. \textbf{08}, 039 (2018).

\bibitem{Harko/2018} T. Harko et al., Phys. Rev. D \textbf{98}, 084043 (2018).



\bibitem{Capozziello/2015} S. Capozziello, V.F. Cardone, and A. Troisi, Phys. Rev. D \textbf{71}, 043503 (2005).
\bibitem{Nojiri/2006} S. Nojiri, S.D. Odintsov, Phys. Rev. D \textbf{74}, 086005 (2006).
\bibitem{Nojiri/2007} S. Nojiri, S.D. Odintsov, J. Phys. Conf. Ser. \textbf{66}, 012005 (2007).

\bibitem{Cai/2012} Y.-F. Cai, D.A. Easson, R. Brandenberger, J. Cosmol. Astropart. Phys. \textbf{1208}, 020 (2012).

\bibitem{Bamba/2014} K. Bamba et al., J. Cosmol. Astropart. Phys. \textbf{01}, 008 (2014).

\bibitem{Nojiri/2016}  S. Nojiri, S.D. Odintsov, V.K. Oikonomou, Phys. Rev. D \textbf{93}(8), 084050
(2016).
 \bibitem{Cai/2014} Y.F. Cai, Sci. China Phys. Mech. Astron. \textbf{57}, 1414--1430 (2014).










\bibitem{Koehn/2014} M. Koehn, J. Lehners, and B.A. Ovrut, Phys. Rev. D \textbf{90}(2), 025005 (2014).
\bibitem{Oikonomou/2015} V.K. Oikonomou, Astrophys. Space Sci. \textbf{359}(1), 30 (2015).

\bibitem{Tolman} R. Tolman, \textit{Relativity, Thermodyanamics and Cosmology} (Dover publications, New York, 1934)

\bibitem{Steinhardt/2002} P. J. Steinhardt and N. Turok, Phys. Rev. D \textbf{65}, 126003 (2002).

\bibitem{Khoury/2004} J. Khoury, P. J. Steinhardt, and N. Turok, Phys. Rev. Lett. \textbf{92}, 031302 (2004).

\bibitem{Singh/2006} P. Singh, K. Vandersloot, and G.V. Vereshchagin, Phys. Rev. D \textbf{74}, 043510 (2006).

\bibitem{Ewing/2013} E. Wilson-Ewing,  J. Cosmol. Astropart. Phys. \textbf{1303}, 026 (2013).

\bibitem{Matter1} J. Quintin, Y. F. Cai, and R. H. Brandenberger, Phys. Rev. D \textbf{90}, 063507 (2014).
\bibitem{Matter2}  Y. F. Cai, R. Brandenberger, and X. Zhang, J. Cosmol. Astropart. Phys. \textbf{03}, 003 (2011).
\bibitem{Matter3} J. de Haro, J. Cosmol. Astropart. Phys. \textbf{11}, 037 (2012).
\bibitem{Matter4}  J. de Haro, Europhys. Lett. \textbf{107}, 29001 (2014).


\bibitem{Brandenberger} R. H. Brandenberger, arXiv:1206.4196.

\bibitem{Ade/2014} P. A. R. Ade et al. (Planck Collaboration), Astron. Astrophys. \textbf{571}, A22 (2014).
\bibitem{Ade/2016} P. A. R. Ade et al. (Planck Collaboration), Astron. Astrophys. \textbf{594}, A20 (2016).

\bibitem{Cai/2009} Y.F. Cai, T. Qiu, R. Brandenberger, X. Zhang, Phys. Rev. D \textbf{80}, 023511 (2009).
\bibitem{Cai/2013} Y.F. Cai, E. McDonough, F. Duplessis, R. Brandenberger, J. Cosmol. Astropart. Phys. \textbf{10}, 024 (2013).
\bibitem{Quintin/2014} J. Quintin, Y.F. Cai, R.H. Brandenberger, Phys. Rev. D \textbf{90}, 063507 (2014).
\bibitem{Haro/2015} J. de Haro, Y.F. Cai, Gen. Relativ. Gravit. \textbf{47}, 95 (2015).

\bibitem{Nojiri/2005} S. Nojiri, S. D. Odintsov, and S. Tsujikawa, Phys. Rev. D \textbf{71}, 063004 (2005).
\bibitem{EoS}  S. Nojiri, S. D. Odintsov, and S. Tsujikawa, Phys. Rev. D \textbf{71}, 063004 (2005); M. P. Dabrowski, arXiv:1407.4851; M. P. Dabrowski and T. Denkieiwcz, Phys. Rev. D \textbf{79}, 063521 (2009).
\bibitem{Odintsov/2015} S. D. Odintsov, and V. K. Oikonomou, Phys. Rev. D \textbf{92}, 024016 (2015).

\bibitem{cmbp} Mustapha Ishak, Amol Upadhye, David N. Spergel, Phys. Rev. D \textbf{74}, 043513 (2006); Shaun A. Thomas et al, J. Cosmol. Astropart. Phys. \textbf{03},  036 (2011); Shant Baghram and Sohrab Rahvar, J. Cosmol. Astropart. Phys. \textbf{12}, 008 (2010);  Jason Dossett et al, J. Cosmol. Astropart. Phys. \textbf{03}, 046 (2014).







\end{thebibliography}
\end{document}